\documentclass[10pt,draftcls,journal,onecolumn]{IEEEtran}

\usepackage{cite}
\usepackage{graphicx}
\usepackage{caption}
\usepackage[cmex10]{amsmath}
\usepackage{amssymb}
\usepackage{amsfonts}
\usepackage{tabularx}
\usepackage{multirow}
\usepackage{bigstrut}
\usepackage{booktabs}
\usepackage{makecell}
\usepackage{algorithm}
\usepackage{algorithmic}
\usepackage[caption=false,font=footnotesize]{subfig}
\usepackage{color}

\usepackage{mdwmath}
\usepackage{mdwtab}
\usepackage{eqparbox}
\usepackage{array}

\begin{document}

\title{Byzantine Attack and Defense in Cognitive Radio Networks: A Survey}

\author{Linyuan Zhang, Guoru Ding,~\IEEEmembership{Member,~IEEE}, Qihui Wu,~\IEEEmembership{Senior~Member,~IEEE,}\\
Yulong Zou,~\IEEEmembership{Senior~Member,~IEEE}, Zhu Han,~\IEEEmembership{Fellow,~IEEE}, and Jinlong Wang,~\IEEEmembership{Senior~Member,~IEEE}

\thanks{This work is supported by the National Natural Science Foundation of China (Grant No. 61172062, 61301160).}

\thanks{L. Zhang, G. Ding, Q. Wu, and J. Wang are with the College of Communications Engineering, PLA University of Science and Technology, Nanjing, Jiangsu, China (e-mail:~zhanglinyuan5@163.com,~dingguoru@gmail.com, wqhtxdk@aliyun.com, wjl543@sina.com). G. Ding and Q. Wu are the corresponding authors.}

\thanks{Y. Zou is with the School of Telecommunications and Information Engineering, Nanjing University of Posts and Telecommunications, Nanjing 210003, China (e-mail: Yulong.Zou@njupt.edu.cn).}

\thanks{Z. Han is with the Department of Electrical and Computer Engineering, University of Houston, Houston, TX, USA (email: zhan2@uh.edu).}}

\maketitle

%Byzantine attack is a classical security problem, while in cooperative spectrum sensing, it is invested with an enriched meaning and novel practical scenarios. Undoubtedly,

\begin{abstract}
The Byzantine attack in cooperative spectrum sensing (CSS), also known as the spectrum sensing data falsification (SSDF) attack in the literature, is one of the key adversaries to the success of cognitive radio networks (CRNs). In the past couple of years, the research on the Byzantine attack and defense strategies has gained worldwide increasing attention. In this paper, we provide a comprehensive survey and tutorial on the recent advances in the Byzantine attack and defense for CSS in CRNs. Specifically, we first briefly present the preliminaries of CSS for general readers, including signal detection techniques, hypothesis testing, and data fusion. Second, we analyze the spear-and-shield relation between Byzantine attack and defense from three aspects: the vulnerability of CSS to attack, the obstacles in CSS to defense, and the games between attack and defense. Then, we propose a taxonomy of the existing Byzantine attack behaviors and elaborate on the corresponding attack parameters, which determine where, who, how, and when to launch attacks. Next, from the perspectives of homogeneous or heterogeneous scenarios, we classify the existing defense algorithms, and provide an in-depth tutorial on the state-of-the-art Byzantine defense schemes, commonly known as robust or secure CSS in the literature. Furthermore, we highlight the unsolved research challenges and depict the future research directions.
\end{abstract}


%\IEEEpeerreviewmaketitle
\newpage

\section{Introduction}
Due to the openness of wireless channels, wireless networks are prone to suffer from various kinds of security threats or attacks, such as jamming, spoofing, and wiretap, etc \cite {wireless_security}. Generally, attackers aim to deteriorate performance of wireless networks in terms of confidentiality, integrity, availability, and access control \cite {S_Frankel}. All these security threats do exist in cognitive radio networks (CRNs), which open a promising paradigm to improve radio spectrum utilization by allowing cognitive or unlicensed secondary users (SUs) to intelligently employ spectrum holes unused by licensed primary users (PUs)~\cite{Haykin_2005}. Apart from the well-known traditional security threats, the cognitive or intelligent nature has introduced into CRNs several new kinds of attacks~\cite{Proc_security,Security-survey,survey-security,review-security,SPM-Varshney}, which block the applications of cognitive radio techniques in commerce and military to a remarkable extent \cite {Burbank_J_L}.

Specifically, spectrum sensing attack is one of the key adversaries to the success of CRNs. According to~\cite{IFA}, spectrum sensing is known as one key enabling technique for cognitive radios to learn from the environment and identify when and where the spectrum holes exist. Nevertheless, in the open wireless environment, spectrum sensing may be misled by malicious adversaries and vulnerable to spectrum sensing attacks~\cite{Clancy_T_C}.

Simultaneously, defense schemes or countermeasures to security threats in traditional wireless networks don't work well when they are confronted with the new spectrum sensing attacks. In non-cognitive wireless networks, countermeasures mainly focus on increasing the signal's robustness to attacks, by reaching agreements between the transmitters and the receivers, such as encryption, authorization and authentication~\cite{wireless_threats}. However, in CRNs, the SU system requires no significant modifications to the PU system \cite {FCC}. The PU system (i.e., detection objective or transmitter) and the SU system (i.e., detector or receiver) are generally separated without signaling exchange. It is mainly the isolation of the PU system and the SU system which makes the countermeasures adapted in traditional networks invalid in CRNs and acts as a major limited condition in designing effective defense schemes against spectrum sensing attacks.

% among which spectrum sensing data falsification (SSDF) attack or Byzantine attack~\cite{SPM-Varshney} has gained worldwide increasing attention during the past couple of years.

Focusing on these new security threats and challenges, novel analysis on attack and defense strategies has gained increasing attention in the past couple of years. Notably, as the focus of this paper, the research on the Byzantine attack and defense has gained significant achievements recently. The Byzantine attack~\cite{SPM-Varshney}, also named the spectrum sensing data falsification (SSDF) attack~\cite{mobile_ad_hoc}, is a kind of insider attack in PHY layer and occurs in the process of cooperative spectrum sensing (CSS)~\cite{IFA}. CSS provides a promising solution to improve the detection performance by exploiting the spatial location diversity of multiple SUs; however, due to the openness of low-layer protocol stacks, CSS is vulnerable to endure Byzantine attackers, who pose serious damage on the reliability of CSS through falsification of the true sensing results.

The main goals of Byzantine attackers in CSS are twofold: the first is to decrease the detection probability for disturbing the normal operation of PU systems, and the second is to increase the probability of false alarm with the purpose of depriving access opportunities of the honest SUs~\cite{kernel_learning}. Proper protection of PU systems is the premise of CRNs, but it is threatened by miss detection. Moreover, data falsification makes the situation worse and even uncontrollable, which poses a great obstacle to the coexistence of CRNs and PU networks. On the other side, the Byzantine attack arises especially when radio spectrum resource is scarce. Through falsifying their sensing outputs and increasing false alarms, malicious users obtain more opportunities of their own communications at the cost of others' loss.

The Byzantine attack problem is not only a threat in CRNs in terms of data security, but also a game between malicious users and the detection system. On one side, false sensing outputs from malicious alteration ruin the integrity of data and lower the reliability of CSS. On the other side, like spear and shield, the Byzantine attack and defense are mutually exclusive but indivisible. When launching the attack, the attackers tend to maximize their attack gains, at the same time of considering how to escape from being found by the defense system; while in turn the diversity and flexibility of attack strategies force the defense system to enhance its universality and efficiency. All of these make the issue of Byzantine attack and defense very challenging and quite complicated. In this paper, we provide a comprehensive survey and tutorial on the recent advances in the Byzantine attack and defense for CSS in CRNs. The main contributions of this paper can be summarized as follows:
 \begin{itemize}
   \item Analyze the spear-and-shield relation between the Byzantine attack and defense from three aspects: the vulnerability of CSS to attack, the obstacles in CSS to defense, and the games between attack and defense.
   \item Propose a taxonomy of the existing Byzantine attack behaviors and elaborate on the corresponding attack parameters, which determine where, who, how, and when to launch attacks.
   \item Make a classification of existing defense algorithms to typical Byzantine attacks and provide an in-depth tutorial on the state-of-the-art Byzantine defense schemes from the perspectives of homogeneous and heterogeneous sensing scenarios, respectively.
   \item Highlight the unsolved research challenges and present the future research directions.
 \end{itemize}

Notably, some similar surveys on security problems in CRNs can be found in the literature \cite{Proc_security,Security-survey,survey-security,review-security,SPM-Varshney}. Specifically, in \cite{Proc_security}, the authors examine both short- and long-term effects of various security threats in CRNs, as well as means of combating them. In \cite{survey-security}, the authors provide a survey of security threats in CRNs that are related to two fundamental characteristics of cognitive radios: cognitive capability and reconfigurability. In \cite{Security-survey}, the authors present a survey of kinds of security aspects in both software defined radio (SDR) and cognitive radio (CR). Compared with the work in \cite{Proc_security,Security-survey,survey-security} that covers much boarder topics and presents surveys from a higher level, this paper focuses on a more specified security aspect, i.e., Byzantine attack and defense in CRNs, and provides a more comprehensive and in-depth survey on this concentrated topic. Furthermore, we find that in \cite{review-security}, the authors make a review of secure CSS in CRNs, and several existing defense algorithms are discussed and compared qualitatively. More recently, the authors in \cite{SPM-Varshney} present the state-of-the-art review on data falsification attacks in a variety of applications in cybersecurity, mainly including distributed spectrum sensing in CRNs and system state estimation in smart grid. Differently, this paper provides a comprehensive survey and holds a mainline to highlight the spear-and-shield relation between the Byzantine attack and defense.

The reminder of this paper is organized as follows. In Section II, we present preliminaries of CSS. In Section III, we analyze the spear-and-shield relation between Byzantine attack and defense. In Section IV, we survey the existing Byzantine attack behaviors. In Section V, we make a tutorial on the advances in Byzantine defense schemes. In Section VI, we present future research directions, followed by the conclusion in Section VII.

\section{Preliminaries of Cooperative Spectrum Sensing}
\begin{figure}[!t]
\centering
\includegraphics[width=0.7\linewidth]{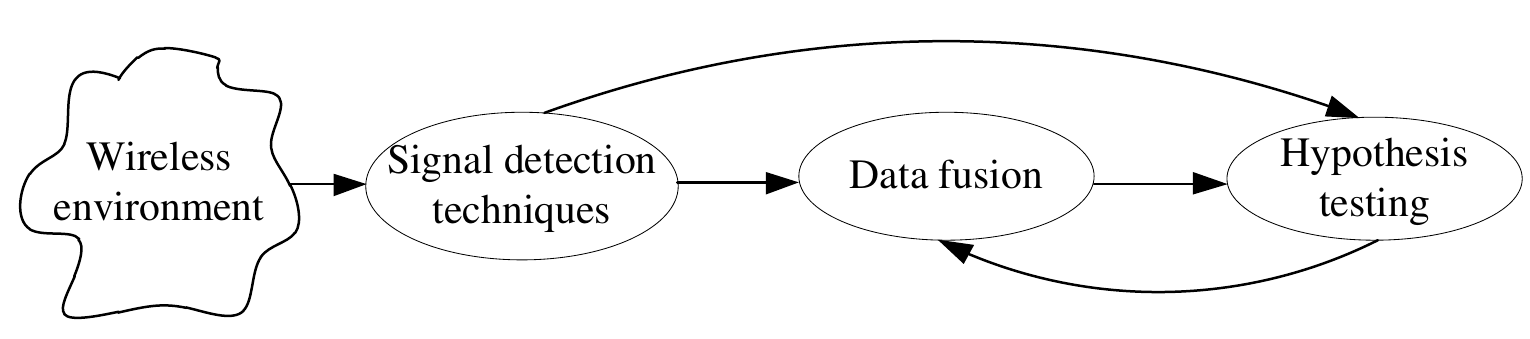}
\caption{Key components of CSS.}
\label{fig_spectrum_sensing}
\end{figure}

Spectrum sensing is used to determine whether a licensed band is occupied or idle, which can be regarded as a binary hypotheses testing problem as follows~\cite{TWC2012}:
\begin{equation}
\left\{ \begin{gathered}
  {H_0}:r(t) = n(t); \hfill \\
  {H_1}:r(t) = h(t)\cdot{P_0}(t) + n(t), \hfill \\
\end{gathered}  \right.
\end{equation}
where the null hypothesis $H_0$ means the absence of the primary signal; while the alternative hypothesis $H_1$ represents the presence of the primary signal. The received signal at time $t$ is $r(t)$, and the noise in the receiver is $n(t)$. ${P_0}(t)$ is the PU's transmit power, and $h(t)$ is the channel gain from the PU to the SU. The uncertainty of channel gains, which integrate the effects of the random noise, path-loss, shadowing, and multi-path fading, makes spectrum sensing of a single sensor unreliable, especially in a low signal-to-noise-ratio (SNR) environment. Consequently, collaborative or cooperative spectrum sensing (CSS) is introduced as a promising method to improve the spectrum sensing performance by exploiting the diversity of multiple SUs. To provide the necessary background for general readers, in this section, three key components of CSS, as depicted in Fig. \ref{fig_spectrum_sensing}, will be briefly discussed:
\begin{itemize}
  \item \emph{Signal detection techniques} are exploited to collect PU's signal on licensed channels and provide the data basis for spectrum sensing. Hence, the selection of signal detection techniques affects CSS's implementation and performances, such as detection accuracy, complexity, and feasibility.
  \item \emph{Hypothesis testing} is a statistic inference method to decide whether the primary users are present or not based on sensing results from signal detection techniques.
  \item \emph{Data fusion} is introduced to combine observations or decisions from multiple SUs to reach a global decision about the availability of the licensed frequency band.
\end{itemize}

\subsection{Signal Detection Techniques}
Among many others, the three popular signal detection techniques for spectrum sensing include: matched filtering detection technique, feature detection technique, and energy detection technique~\cite{sensing_technique}. Compared to the other two methods, the energy detection cannot differentiate noises from primary users' signals, but it does not need any prior knowledge about the PU signals. Moreover, among the sensing technologies, energy detection is of low complexity. Specifically, let $r(t)$ denote the $t$-th sample of the received signal and $y$ denote the output of energy detection. The decision rule can be formulated as follows \cite  {energy_detection}:
\begin{equation}
\label{local_test}
y = \sum\limits_{t = 1}^N {{{\left| {r(t)} \right|}^2}} \mathop  \gtreqless \limits_{D = {H_0}}^{D = {H_1}} \eta,
\end{equation}
where $N$ is the time-bandwidth product, $D$ denotes the decision result, and $\eta$ is the decision threshold. Two vital metrics for evaluating the detection performance are: the probability of false alarm $P_f$ and the probability of miss detection $P_d$, which can be defined as ${P_f} = \Pr(D={H_1}|{H_0})$ and ${P_m} = \Pr(D={H_0}|{H_1})$, respectively. Generally speaking, low false alarm probabilities and high detection probabilities are desired for CRNs. False alarms prevent spectrum holes from being accessed by secondary users, and thus result in spectrum under-utilization; while correct detections of the primary signals' presence avoid harmful interference to primary users.

\subsection{Hypothesis Testing}
Hypothesis testing is performed to obtain the binary decision on the presence of the primary signal. There are different testing methods adopting various design rules. Here we investigate three widely adopted hypothesis testing methods, i.e., the Bayesian test, the Neyman-Pearson test, and the sequence probability ratio test, in the sequel~\cite{hypotheses}.

% brought about by false alarms and miss detections. The a prior probabilities of the two hypothesis are needed, denoted by $P(H_0),P(H_1)$ respectively. The Bayesian test can be given as
\subsubsection{Bayesian Test}
The Bayesian test is proposed to minimize the average detection cost or risk given by
\begin{align}
C_{\rm{average}} = \sum_{i\in\{0,1\}}\sum_{j\in\{0,1\}} C_{i,j} P(H_j) \Pr(u=H_i|H_j),
\end{align}
where $P(H_i),i\in\{0,1\}$ denotes the prior probabilities of the underlying hypothesis, $\Pr(u=H_i|H_j), i,j\in\{0,1\}$ is the conditional probability of declaring $u=H_i$ when $H_j$ is true, and $C_{i,j}, i,j\in\{0,1\}$  is the corresponding cost. To minimize the average detection cost, and the optimum decision rule is derived as~\cite{Optimal_1986}:
\begin{equation}
T_{\rm{Bayesian}} = \frac{{p(u|{H_1})}}{{p(u|{H_0})}} \mathop  \gtreqless \limits_{D={H_0}}^{D={H_1}} \frac{{P({H_0})({C_{10}} - {C_{00}})}}{{P({H_1})({C_{01}} - {C_{11}})}}.
\end{equation}

\subsubsection{Neyman-Pearson Test}
In many practical situations, not only are the prior probabilities unknown, but also the cost assignments are difficult to make. In this case, the Neyman-Pearson (NP) test is introduced to minimize the probability of miss detection, $P_m$, while maintaining the probability of false alarm $P_f$ to be lower than an acceptable value $\alpha$, i.e.,
\begin{align}
\min ~~P_m~~~~~~
{\rm{s.t.}} ~~P_f \le \alpha.
\end{align}
With simple mathematical derivations, the resulting test is performed as follows~\cite{NP-1993}:
\begin{equation}
  T_{\rm{NP}} = \frac{{p(u|{H_1})}}{{p(u|{H_0})}}\mathop  \gtreqless \limits_{D = {H_0}}^{D = {H_1}} \lambda,
\end{equation}
where $\lambda$  is the detection threshold calculated based on the maximum acceptable probability of false alarm.

\subsubsection{Sequence Probability Ratio Test}
Different from the above two hypothesis testing methods of a fixed-sample-size, the sequence probability ratio test (SPRT) is proposed to shorten the sensing time, where samples are collected sequentially and a final decision is made as soon as the test statistics $T_{\rm{SPRT}}(k)$, from $k$ samples, satisfies the decision quality~\cite{SPRT}. Specifically, the test procedure can be represented in such a way~\cite{SPRT}:
\begin{equation}
\left\{ \begin{gathered}
  T_{\rm{SPRT}}(k) \leqslant {\eta _0} \Rightarrow D={H_0}, \hfill \\
  T_{\rm{SPRT}}(k) \geqslant {\eta _1} \Rightarrow D={H_1}, \hfill \\
  {\eta _0} < T_{\rm{SPRT}}(k) < {\eta _1} \Rightarrow \rm{take~another~sample}, \hfill \\
\end{gathered}  \right.
\end{equation}
where ${\eta _0}$, and ${\eta _1}$ are two detection thresholds calculated by sensing requirements, and ${\eta _0} < {\eta _1}$. New observations are taken only when the current sensing results are not sufficient. The main advantage of SPRT is that it requires, on the average, fewer samples to achieve the same detection performance as the fixed-sample-size testing methods~\cite{hypotheses}.

\subsection{Data Fusion}
In CSS, each SU firstly performs local spectrum sensing so as to obtain its individual sensing output. Then, data fusion is made to combine the sensing outputs from multiple SUs and reach a more reliable sensing output by exploiting the spatial diversity. Note that data fusion rules keep robustness to attack and the choice of fusion rules deserves considerable attentions \cite{Cai_Y,double_threshold}. Based on the network infrastructure, data fusion in CSS can be classified into two categories: centralized fusion and decentralized fusion.

\subsubsection{Centralized Fusion}
In centralized networks, SUs individually perform local sensing and report the sensing results to the dedicated node called fusion center (FC). Then, the FC fuses the received sensing results and determines the state of the primary signal. Obviously, the selection of fusion rules is important~\cite{Cai_Y}. The local sensing reports have different types, such as binary sensing results, soft observations, and quantized values, which correspond to the data fusion in terms of hard~\cite{hard_fusion}, soft~\cite{optimal_fusion}, and quantized fusion \cite{bi_weight}.

\subsubsection{Decentralized Fusion}

Unlike centralized fusion which relies on a FC to collect the data and make the fusion, for decentralized fusion in cognitive radio ad-hoc networks (CRAHN)~\cite{CRAHN}, no dedicated central entity exists to perform data fusion, and instead, each sensor exchanges its sensing output with its neighbors and iteratively fuses the sensing outputs from its neighbors. In the literature, representative algorithms for decentralized fusion include consensus algorithms~\cite{Proc-Consensus,TVT-Consensus,TSP-Consensus,ICC2012}, diffusion algorithms~\cite{diffusion} and belief propagation algorithms~\cite{TWC-BP,TWC-BP2,JSAC-BP}, etc.

\section{Attack and Defense: Spear and Shield}
%Firstly, spectrum sensing may be interrupted by Primary User Emulation (PUE) attack in which an insider or outsider mimics a primary user signal's characteristics, thereby causing legitimate secondary users to mistakenly declare the presence of PU and then switch to another channel. Secondly, the common control channel used to report local sensing results may be interfered with by Denied-of-service (DoS) attackers through launching jamming attack in which attackers release energy on the channel, increase bit error probability and cause the transition failure of measurements.
%Apart from these attacks, a more stealthy but destructive attack exists inside of the network.

\subsection{Vulnerability of CSS to Attack}	

The openness of both wireless channels and software-defined radio (SDR) platforms increases the risks of CSS, which provides malicious users with plenty of opportunities to access CRNs and disguise as normal sensing nodes. After intrusion, malicious users report falsified results instead of real observations to disturb normal performances of both primary and secondary networks. Moreover, selfishness of sensing nodes motivates them to report falsified sensing results and increase their own gains at the cost of other honest users' performance degradation. Typically, malicious users intentionally conduct the spectrum sensing data falsification (SSDF) attack or the Byzantine attack in pursuit of two attack objectives as follows \cite{classification,Fatemieh_O}:
\begin{itemize}
  \item \emph{Vandalism objective}: Interference to the primary system. This class of malicious users report channel vacancy when the sensing results indicate that the channel is busy. The malicious action induces the FC to make more miss detection mistakes, allowing of sensors' false access to the busy channel, and imposes severe interference on the PUs. Interference to the primary system violates the basic premise of CRNs, and then discourages PUs from sharing their licensed spectrum bands with SUs.
  \item \emph{Exploitation objective}: Exclusion of idle channels. This class of malicious users send channel busy information when their sensing results indicate that the channel is idle. In consequence, the FC falsely declares the channel busy and has to order other SUs to wait for another sensing period or switch to other channels. Then the malicious users can access to the idle channel exclusively, at the expense of other SUs.
\end{itemize}
 	
Although the two objectives above are listed separately, it does not mean that a malicious user must conduct attacks for only one objective. In fact, malicious users can attack for both two objectives and be in pursuit of attack utility maximization~\cite{IEEE-TCOM-Ding}.

\subsection{Obstacles to Defense}
\begin{figure}[!t]
\centering
\includegraphics[width=0.9\linewidth]{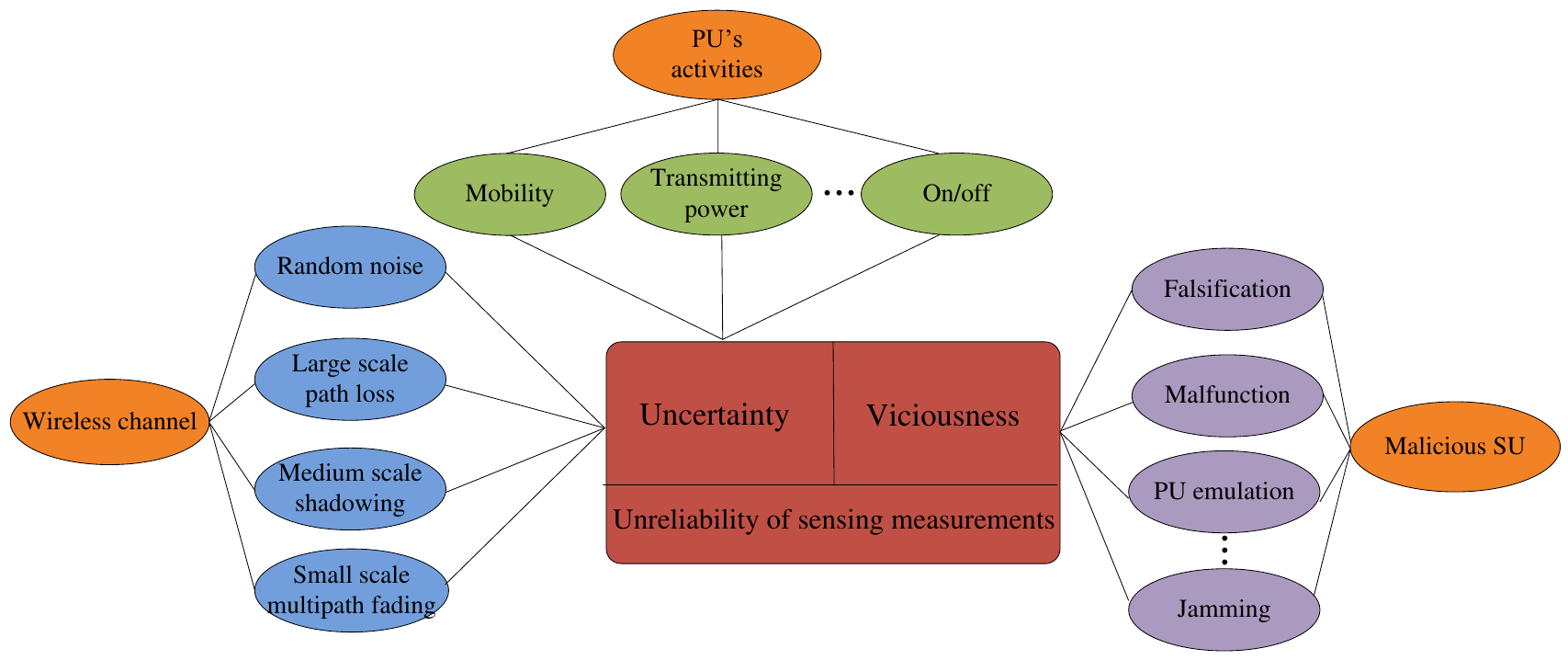}
\caption{Sources of unreliability in cooperative spectrum sensing.}
\label{fig_unreliability}
\end{figure}

Before talking about defense strategies counteracting the Byzantine attack, we need to understand the existing obstacles or challenges. One critical obstacle is to make correct distinction between malicious users and legitimate (or honest) ones in CSS. As shown in Fig. \ref{fig_unreliability}, the process of CSS is closely related to PUs' activities, SUs' behaviors and wireless channels' characteristics, all of which contribute the sources of the unreliability of spectrum sensing measurements in CSS. Uncertainty means that even legitimate SUs may report false values due to limited sensing capability or effects of uncertain sensing environment. Viciousness means that a SU is essentially malicious and intentionally reports falsified values. Both two factors contribute to the unreliability of measurements, deteriorating performance of CSS. The SU network does not have the exact acknowledge of PUs' state and has to rely on the unreliable measurements to perform CSS.
Specifically, uncertainty and viciousness are similar in term of their effects on reports. One SU's reports become erroneous at some times, and are inconsistent with others'. From this perspective, uncertainty of measurements helps malicious users hide from being detected.

Most papers focus on two problems: uncertainty and viciousness of measurements. Undoubtedly, uncertainty of observations is an essential attribute of spectrum
sensing and can be regarded as one resource of unreliability of measurements. At the same time, malicious users inject falsified sensing results representing viciousness of measurements and exacerbate the unreliability in a similar manner. That is, when there exist malicious users in the SU network, uncertainty and viciousness give rise to the unreliability of sensing measurements and must be tackled with together.

\subsection{Game between the Byzantine Attack and Defense}

Naturally, the Byzantine attack greatly degrades CSS through reporting fake sensing results, while the Byzantine defense aims to eliminate negative effects of attacks, achieve effective exploitation of idle spectrum resources, and keep harmonic with the PU network.

\begin{figure}[!t]
\centering
\includegraphics[width=0.6\linewidth]{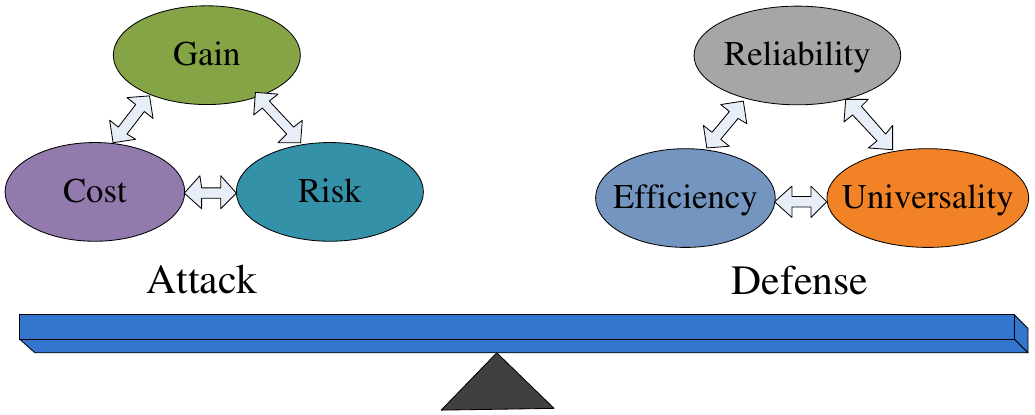}
\caption{Game between Byzantine attack and defense.}
\label{fig_tradeoff}
\end{figure}

To investigate the issue of the Byzantine attack and defense, there is a fundamental problem: What are the best strategies for attackers and defenders, respectively maximizing their objectives \cite {Rawat_A_S}\cite {distributed_bayesian}. Indeed, there is a game between the Byzantine attack and defense.
Their relation is illustrated in Fig. \ref{fig_tradeoff}. On one hand, attack and defense are mutually opposite. On the other hand, there are several interesting tradeoffs in the Byzantine attack and defense, respectively.

From the perspective of Byzantine attackers, when launch the attack, Byzantine users have to take into account the factors: attack gain, attack cost and attack risk, explained as follows:
\begin{itemize}
  \item \emph{Attack gain} denotes exploitation gain and vandalism gain obtained through falsification;
  \item \emph{Attack cost} includes intrusion cost, communication cost, and so on, which are expended to launch attack;
  \item \emph{Attack risk} can be formulated as the product of the probability of being identified and consequent potential punishment.
\end{itemize}

Attack gain's persistence is influenced by attack risk, i.e., malicious users have to endure more attack risk if they are in pursuit of extra attack gains~\cite{paper}; on the other hand, increased attack cost can help enhance attack gain and lower attack risk. For example, coordinated attack among malicious users improves the probability of the final decision being dominated, but they have to pay for the extra communication cost during the coordination process.

Similarly, from the perspective of defense schemes, there are also three aspects deserving consideration:
\begin{itemize}
  \item \emph{Defense reliability}, which is closely related to attack situations and the final spectrum sensing performance together. For example, the maximum acceptable malicious users under limitation of detection performance can be used to evaluate the effectiveness.
  \item \emph{Defense efficiency}, which includes convergence rate \cite{enhance_wsprt} \cite{Soltanmohammadi}, defense cost \cite {Althunibat}, and computation complexity. Convergence rate is an essential metric and the lower the convergence rate is, the shorter effective time for transmitting is, especially in decentralized networks~\cite{Yan_q}.
  \item \emph{Defense universality}, which characterizes the universality of defense algorithms. Diversity of attack strategies unknown by defenders poses a severe challenge on the universal design of the defense scheme.
\end{itemize}

Majority of existing defense algorithms focus on improving defense reliability to specific attack behavior. In consequence, defense efficiency and universality are generally ignored. As stated in~\cite{Rawat_A_S}, defenders have to employ different defense schemes for different attack situations, which sacrifice the defense universality. In~\cite{IEEE-TCOM-Ding}, a generalized modeling approach for sensing data with an arbitrary abnormal component is proposed, and a data cleansing-based robust spectrum sensing algorithm is developed to tackle this generalized attack.

In a nutshell, Byzantine attack and defense are mutually restrained from each other. Attackers need to adjust their strategies to keep their effects on final decisions and avoid defenders' detection, while defenders have to learn and analyze attack behaviors and designs effective defense rules. Indeed, attack and defense ought to be considered together, no matter whether it is from the perspectives of attackers or defenders. Attackers have to hide themselves from being detected in order to optimize their attack performance, while defenders have to find out the clues to attack behaviors.

\section{Byzantine Attack}

To provide a comprehensive and deep understanding on the Byzantine attack, this section will answer about where, who, how, and when to launch attacks and highlight several typical attack models.

\subsection{Attack Parameters}
\begin{figure}[t]
\centering
\includegraphics[width=0.45\linewidth]{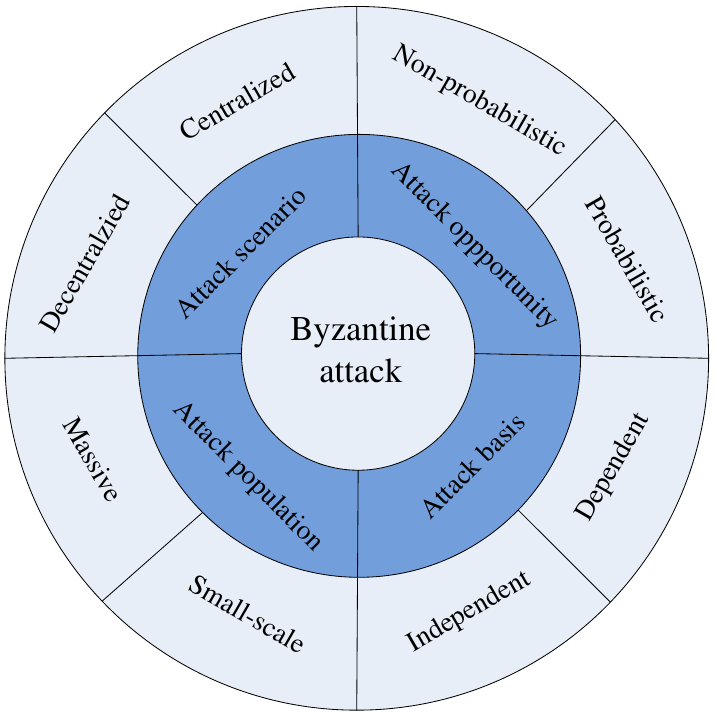}
\caption{A taxonomy of Byzantine attack parameters.}
\label{fig_attack}
\end{figure}

An essential characteristic of the Byzantine attack is its flexibility and diversity, which pose great challenges on detection of malicious users. As shown in Fig.~\ref{fig_attack}, the Byzantine attack can be described mainly by four parameters, attack scenario, attack population, attack basis, and attack opportunity, which answer about where, who, how, and when to launch the Byzantine attack, respectively. The four parameters are briefly explained as follows:
\begin{itemize}
  \item \emph{Attack scenario} depicts where attackers act in, i.e., the network circumstance. CSS is classified into centralized and decentralized based on whether there exists a node receiving and fusing all sensing results. The type of CSS determines the way of SUs cooperating, greatly influencing attack behaviors, and is chosen as the classification standard of networks.
  \item \emph{Attack basis} denotes information based on which malicious users falsify measurements and solves how to launch attacks. The original information is local sensing results, and extra information includes other SUs' sensing results, fusion rules, and so on. An abundant attack basis
      can assist malicious users to design effective attack strategies.
  \item \emph{Attack opportunity} answers about when to attack. At a certain slot, malicious users may decide whether or not to attack with a certain probability. Besides, attackers may take into consideration of this current state, such as sensing results and attack expectation \cite{catch_me}. Attack opportunity reveals the flexibility and variety of attack behaviors.
  \item \emph{Attack population} is a basic metric of the Byzantine attack answering about who to attack, and is equal to the percentage of malicious users in all secondary users, dedicating to what degree of severity networks suffer from attackers. Naturally, as the attack population increases, results of data fusion leans more to the Byzantine users.
\end{itemize}

\subsection{Attack Scenario: Where to Attack?}
An attack scenario is not an attacker's own attribute but is the circumstance attackers have to adapt to. In other words, where to attack determines what kind of network drawbacks can be taken advantage of to launch powerful attacks by malicious users.

Previous studies focus on SSDF attacks in centralized CSS \cite {WSPRT}\cite {Bhargava_V_K} where it is always assumed that a reliable FC exists for receiving and processing all measurements. Although fake data from malicious users and real data from legitimate users are mixed, it is possible to analyze behaviors of both malicious and honest SUs accurately, especially in cases that honest ones are in majority. Hence, the centralized scenario may not be favorable for attackers.

Compared to centralized CSS, decentralized CSS is more vulnerable to SSDF attacks \cite {Yan_q}, because in decentralized CSS, no users can obtain all other users' real value as FC to make decisions, and instead, global decisions are made in the manner of iteration through information exchange with the neighbors. Firstly, observations of legitimate neighbors and the network's detection algorithms are known by malicious users and can be used to assist them to launch attacks. Secondly, fake data can be integrated into neighbor nodes' decisions and stealthily diffused among the network, which eventually brings significant damages on detection performances. Last but not least, the iteration process for reaching a consensus is vulnerable to be exploited by malicious users through continuous injection of vicious data.

\subsection{Attack Basis: How to Attack?}
Attack behavior is established on attack basis, i.e., information obtained by attackers. Specifically, attack basis provides a solution to the question: How to attack. From the perspective of malicious users, accurate attack is attractive, effective, but hard to reach. Generally, malicious users don't have the knowledge of licensed channels' real occupation state. Indeed, a malicious user may just have some sensing capability itself, which is called independent attack. This kind of attack is more or less rough and only some simple attack strategies can be implemented. In consequence, the attack performance is always limited and attackers are vulnerable to be detected.

Nevertheless, in dependent attack, some extra information, beyond its own sensing results, is obtained by an attacker, such as the fusion rule \cite {distributed_bayesian}, defense strategies \cite {hit_and_run}, and current sensing results of other malicious users. The information can improve flexibility of attack strategies and enhance attack intensity. Especially, through communication between malicious users, an attacker can collude with its peers. Before reporting sensing results, malicious users can exchange local sensing results, and launch coordinated attacks. Information exchange makes attacks more accurate and coordinated attacks enhance the power of attack, increasing the success rate of attacks. In addition, the attacker can also improve its sensing accuracy and attack's effectiveness through eavesdropping honest users' reporting. In decentralized networks, malicious nodes' collusion poses more severe security threats, because the collusion of attackers cannot only help to achieve a faster convergence rate to their desired objectives, but also provide them with an opportunity to support each other in the trust verification schemes \cite {Yan_q}. However, information exchange among the Byzantine attackers is not easy due to the multi-hop character of the decentralized CSS, which may accelerate exposure of attack behaviors instead.

%After all, dependent attackers can launch more destructive attacks. Considering the importance of attack basis, we call a kind of attack the independent attack
%when its attack basis is limited to the malicious user's own local information just same with honest one's, and otherwise, the dependent attack.

\subsection{Attack Opportunity: When to Attack?}
The next critical problem is when to attack. Selection of attack opportunity will influence attack gains and risks. If attackers report falsified results all the time, such attackers catch every attack opportunity no matter whether attacks work or not. In consequence, they obtain significant but transient attack performance, as frequent falsification makes them prone to be detected by defense mechanisms. In practice, an attacker may not misbehave all the time, especially if it is in pursuit of long-term profits. So when does a Byzantine user attack? A solution easy to come up with is the probabilistic attack where a malicious user launches attacks with a certain probability \cite {Penna_F}. The probabilistic attack can improve stealthiness of attack behaviors, and is easy to be modeled and analyzed.

However, the probabilistic attack keeps attack behaviors consistent over time slots, and defense algorithms can identify the attackers through statistical analysis. Besides, the probabilistic attack ignores current observations during the process of selection of attack slots. Compared to the probabilistic attack, the non-probabilistic attack, is more elusive but hard to be modeled. Take a simple example, an attacker decides to launch attacks only when its observations are lower than a threshold ${\eta _0}$ or higher than another threshold ${\eta_1}$, ${\eta _1} > {\eta _0}$. Though the attack model is probabilistic from the statistical perspective, it is still non-probabilistic, because it decides whether to attack based on current slots' observations.

\subsection{Attack Population: Who to Attack?}
Attack population answers about who to attack. Specifically, it is equal to the percentage of malicious users in all SUs and indicates to what degree of severity the networks suffer from the Byzantine attack. Naturally, attack performances are improved with increase of attack population. Indeed, once the attack population exceeds a certain value, CSS does not work any more. Hence, we need to exactly formulate the relation between the attack performances and the attack population. In \cite {Rawat_A_S}, a measurement metric, the Kullback-Leibler divergence (KLD), is exploited to characterize detection performance as follows:
\begin{align}
D(P(u|{H_1})||P(u|{H_0}))
= \sum\limits_{u \in \{ 0,1\} } {P(u|{H_1})} \log \frac{{P(u|{H_1})}}{{P(u|{H_0})}},
\label{KLD}
\end{align}
where $u$ is the final decision of the network obtained based on received measurements. KLD reflects on the value of the final decision and a low value of the KLD indicates its neglectable value in determining the primary user's state. Specifically, when the KLD becomes zero, the FC is regarded as blind, because FC's decision is no better than tossing a coin. Here the value of attack population is named blind point. We call attacks with an attack population under the blind point as the massive attack, otherwise the small-scale attack. In addition,  attack population's effects on  attack performance are related to fusion rules to a considerable extent \cite{physical_layer,Abdelhakim_icassp}.

\subsection{Typical Attack Models}
\begin{figure}[h]
\centering
\includegraphics[width=\linewidth]{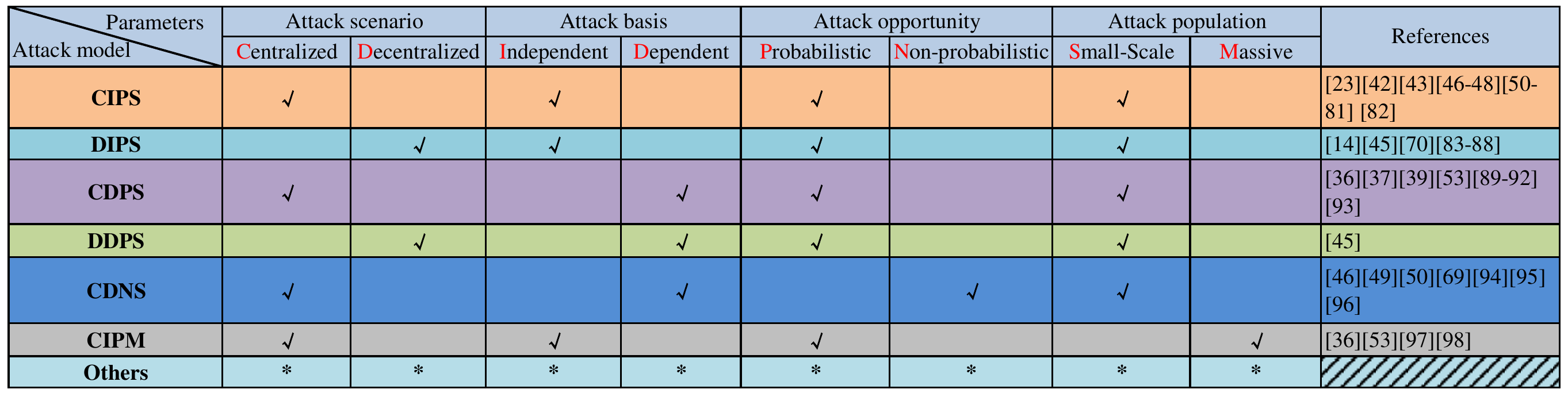}
\caption{Current work on the Byzantine attack, where  attack models are denoted with the combinations of initial letters of their attack parameters, the check character indicates that the corresponding attribute is related, the asterisk character means any choose, and the slash character means that there are no related works currently.}
\label{fig_current_attack}
\end{figure}

In this subsection, we will further discuss some typical attack models, i.e., typical combinations of the four attack parameters, in order to give a comprehensive view of attack behaviors. Fig. \ref{fig_current_attack} outlines current researches on attack models.

\begin{figure}[h]
\centering
\includegraphics[width=0.8\linewidth]{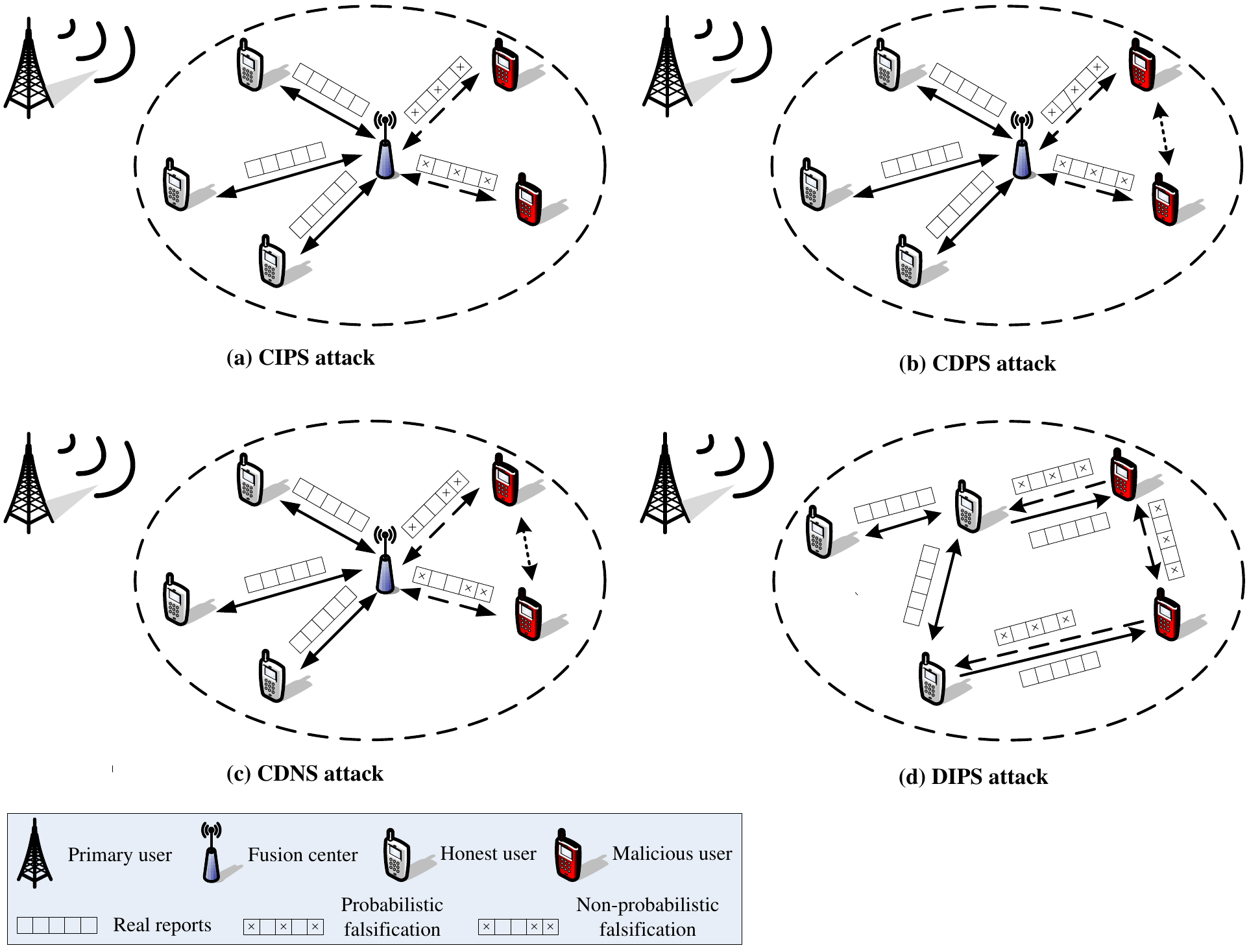}
\caption{Typical attack models.}
\label{fig_typical_attack}
\end{figure}

Based on Fig. \ref{fig_current_attack}, we select four typical attack models as illustrated in Fig. \ref{fig_typical_attack} and provide with detailed description in sequence.

%CIPS: \cite{Feng_DS} \cite {Marano_S},\cite{adaptive_byzantine}\cite {double_threshold}\cite {Nguyen_Thanh}\cite {enhance_wsprt}\cite {Soltanmohammadi}
%\cite {catch_me,WSPRT,Bhargava_V_K}\cite {Penna_F,physical_layer,Abdelhakim_icassp,new_technique,Sun_Y,CatchIt,Min_a_w,shadow_fading,Kaligineedi_P_2010,zhang_x,Pena_F_glo,Believe_yourself,Non-parametric,Mins_a_w,kasiri_B,Attack_prevent,mobile_reputation,Jana_S,HMM_based,dalianli,no_regret,cross_layer,Fastprobe,multinode_sensing,decoupling,opti_malicious,Sun_Y_2010,Verikoukis_C}\cite{Safduyu_milcom}
%
%%\cite{counterattack}
%
%DIPS: \cite {mobile_ad_hoc}\cite {Yan_q}\cite {mobile_reputation} \cite {distributed_chen,reiforcement_learning,ReDiSen,deviation_tolerant}\cite{biologicall_2010}\cite{selfish_2013}
%
%CDPS: \cite {Rawat_A_S} \cite {Marano_S}\cite{optimal_distributed}\cite {classification}\cite {Fatemieh_O}\cite {Zhengrui,Wang_J_2013,abnormality_icc,block_outlier}
%
%DDPS: \cite {Yan_q}
%
%CDNS: \cite {catch_me}\cite {hit_and_run}\cite {Penna_F}\cite {Attack_prevent}\cite{dos,online_yao,intelligent_attack}
%
%CIPM: \cite {Marano_S}\cite {Xiaofan_He}\cite {classification}\cite {Przemysaw_Paw}

\subsubsection{Centralized Independent Probabilistic Small-Scale (CIPS) Attack}
Here, we consider a case in a centralized network, where a small-scale group of attackers randomly and independently falsify measurements with a certain probability,as shown in Fig. \ref{fig_typical_attack} (a). This is denoted as CIPS attack as shown in Table 1. The general CIPS attack model can be formulated as follows:
\begin{equation}
\left\{ \begin{gathered}
  P({X_m^{'}} = {X_m} + \Delta |{X_m} < \eta ) = {\alpha _0}, \hfill \\
  P({X_m^{'}} = {X_m} - \Delta |{X_m} > \eta ) = {\alpha _1}, \hfill \\
\end{gathered}  \right.\begin{array}{*{20}{c}}
  &\rm{exploitation~attack}, \\
  &\rm{vandalism~attack},
\end{array}
\end{equation}
where $X_m$ is the malicious user's observation, $\Delta$ is a positive value called the attack strength, $X_m^{'}$ is falsified sensing results, ${\alpha _0}$ is the attack probability of false alarm, and ${\alpha _1}$  is the attack probability of miss detection, and $\eta$ is attack threshold. This probabilistic attack model is widely applied in related literatures \cite {Rawat_A_S}\cite {catch_me}\cite {Penna_F}. Specifically, $({\alpha _0},{\alpha _1}) = (1,0),(0,1),(1,1)$, respectively, correspond to always yes/no/adverse attack in \cite {WSPRT}\cite {Bhargava_V_K}\cite {Nguyen_Thanh}. In addition, when both probabilities are zero, the user is honest and reports real measurements all the time.

It is important to point out that we choose soft falsification to describe attack behaviors, and hard falsification is similar and easy to be inferred from soft falsification. Obviously, differences between the two falsifications are that instead of introducing $\Delta $, attackers just flip raw decisions in hard reporting. Soft reporting is regarded as a flexible and stealthy way for attackers due to its abundant value space. But soft falsification makes performance analysis in theory more complex. So we take hard falsification as an example and evaluate attack population's effect on attack performances. In \cite {Rawat_A_S}, based on the KLD in (\ref{KLD}), the blind point is formulated as follows:

\begin{equation}
{\alpha _{\rm{blind}}} = \frac{{P_d^H - P_{fa}^H}}{{(P_d^B - P_{fa}^B) + (P_d^H - P_{fa}^H)}},
\end{equation}
where $P_d^H (P_d^B)$ denotes the detection probability of an honest (malicious) SU, and $P_f^H (P_f^B)$ denotes the false alarm probability of a honest (malicious) SU. Obviously, Byzantine users' local sensing capability is vital to attack performance. Specifically, when all users share the same local sensing performance, e.g. all users locate in a small area, which is assumed by most of previous studies, the blind point is 0.5. That is to say, when the number of honest SUs is same with that of malicious SUs, the fusion results show no value for spectrum sensing.

\subsubsection{Centralized Dependent Probabilistic Small-Scale (CDPS) Attack}

Different from the CIPS attack which relies only on local sensing, extra information is exploited in the CDPS attack, as shown in Fig. \ref{fig_typical_attack} (b).
For example, the FC's fusion rule as known prior knowledge can be used to optimize attack strategies \cite {distributed_bayesian}\cite {optimal_distributed}. One kind of CDPS attack is cooperative attack in which malicious users communicate with each other. One direct gain from the cooperative attack is improvement of spectrum sensing capability. The blind point decreases with attackers' sensing capability, meaning that to achieve certain attack performances, the cooperative Byzantine attack needs a smaller population than the independent attack. Besides, information exchange can guide malicious users to effectively design attack strategies. In general, cooperative malicious users   \cite {Fatemieh_O} \cite {Rawat_A_S}\cite {Zhengrui} \cite {Wang_J_2013} firstly exchange measurements with each other, guaranteeing more accurate decision about the licensed channel's state, and then report coordinated falsified results to enhance attack power. Furthermore, another attack model called the omniscient attack \cite {classification} is proposed, where attackers don't only know measurements of all attackers, but also know legitimate users' and the fusion rule. In consequence, omniscient attackers can falsify reports so that final decisions are wrong, as well as decreasing chances of being detected. In other words, when more information is obtained, except of manipulating final decisions, attackers pay increasing attention on improving attack stealthiness. In \cite {Wang_J_2013}, to achieve the two-fold objectives, attack destructiveness and stealthiness, attackers randomly select measurements of honest nodes to report, which deteriorates detection performances through lowering the spatial diversity gain at the same time of hiding attackers from being detected.

Note that in related works, attacks referred to above are "always attack", a special case of probabilistic attack, ignoring attack stealthiness to some extent. Indeed, when the attack probability is appropriately set, dependent attackers can imitate honest users' statistic characteristics, avoid being detected under existence of known trust nodes, and launch cooperative and stealthy attack \cite {abnormality_icc}.

\subsubsection{Centralized Dependent Non-probabilistic Small-Scale (CDNS) Attack}
Compared to the probabilistic attack above, non-probabilistic attackers consider the current state and flexibly adjust the attack behavior, which decreases attack risk and improve attack gains. Here, we take the CDNS attack as an example to focus on illustrating the non-probabilistic characteristic coordinated with other attack characteristics, as shown in Fig. \ref{fig_typical_attack} (c). Especially, dependence plays an important role in the non-probabilistic attack. In \cite {catch_me}, Byzantine nodes are omniscient like attackers in \cite {classification}, but they report one-bit local decision. Under the limitation of attack population, attackers cannot reverse final decisions at some time. Hence, they launch attacks only when they can do. This attack behavior is non-probabilistic as attackers consider the current state, i.e., attack expectancy. In \cite {hit_and_run}, this intention is expressed more obviously. Attackers in \cite {hit_and_run} know reports of all other nodes, so they can calculate the suspect lever as their posteriori probability of being malicious ones \cite {CatchIt}. If the suspect lever is lower than a threshold, attackers continue launching attacks; otherwise, they stop to report real observations until the suspect lever resumes to a secure degree. This kind of non-probabilistic attack considers current situation before making attack decisions so that attackers avoid excessive attacks and saves attack resources. Besides, without extra information, attackers can apply a time-varying attack probability \cite {Penna_F} requiring better reactivity of the secure algorithm.

In addition, there exists a special kind of attackers determining whether or not to attack based on utility function \cite {Attack_prevent}\cite {dos}\cite {intelligent_attack}. Only if falsification of reports increase their utilities, malicious users will launch attacks; and otherwise, no attacks. This kind of intelligent attacker considers the attack behavior from the perspective of attack objectives and avoids meaningless attacks.

\subsubsection{Decentralized Independent Probabilistic Small-Scale (DIPS) Attack}
In some cases, the DIPS attack is similar to the CIPS attack \cite {Believe_yourself}\cite {reiforcement_learning}\cite {ReDiSen}\cite {deviation_tolerant}\cite{biologicall_2010}\cite{selfish_2013}. Nevertheless, due to decentralized CSS's essential characteristics, a user receives measurements from its neighbors and has the knowledge of algorithms for detecting attackers and determining states of licensed channels, which can well assist it to draw up novel attack strategies \cite {Yan_q}, as shown in Fig. \ref{fig_typical_attack} (d). So next we will mainly elaborate on how attackers take advantage of the network's drawbacks to implement much more stealthy and effective attack.

In centralized networks, attack opportunity answers about the selection of slots to attack; while in decentralized networks, attack opportunity does not only involve with the selection of attack slots, but also involves with the choice of iteration times to attack during the convergence process.
That is to say, a malicious user may not only manipulate sensing data in the sensing stage, but also inject fake state values in the iteration stage \cite {deviation_tolerant}.
During the convergence process, users receive updated state value timely from the neighbors and track convergence trend, which enables attackers to keep a good check on its attack performances. At the same time, it can enhance attack performances through continuous injection of fake state values. From this perspective, a kind of attack called the covert adaptive data injection attack is considered in \cite {Yan_q}. Assuming that an outlier detection algorithm using a static attack detection threshold is adopted in the network, a Byzantine user computes the maximal acceptable deviation based on all its neighbors' submitted states and the threshold in the detection algorithm, and then transmits its forged value to its neighbors \cite {Yan_q}.

However, falsification of sensing results, especially fake state values, inevitably thwart the convergence process, finally extending the convergence time and shortening the effective time for communication. There is no benefit for both honest users and malicious ones. Indeed, malicious nodes do not need to inject fake values any more just when the conditions below are satisfied \cite {Yan_q}.
\begin{equation}
\label{coverage}
\begin{gathered}
  \bar x + \frac{{\sum\nolimits_{i = 0}^{{k_{\rm{stop}}}} {a(i)} }}{m} > \gamma ,a(i) > 0, \rm{exploitation~objective}, \hfill \\
  \bar x + \frac{{\sum\nolimits_{i = 0}^{{k_{\rm{stop}}}} {a(i)} }}{m} < \gamma ,a(i) < 0, \rm{vandalism~objective}, \hfill \\
\end{gathered}
\end{equation}
where $\bar x$ is the average value of the original measurements of the whole network, which is also the legitimate consensus result, $m$ is the number of nodes in the network, $\gamma $ is the detection threshold used to determine the state of PU, $a(i)$ is forged value at the $i$-th step, and ${k_{\rm{stop}}}$ denote the final step to inject fake value. Absolutely, metric such as $\bar x$  is hard to obtain but the left part of (\ref{coverage}) can be estimated in \cite {Yan_q}. That is to say, once the attack objective is achieved, the attack behavior is finished so that convergence time is shortened.

\subsection{Discussions}
\begin{figure}[!h]
\centering
\includegraphics[width=1\linewidth]{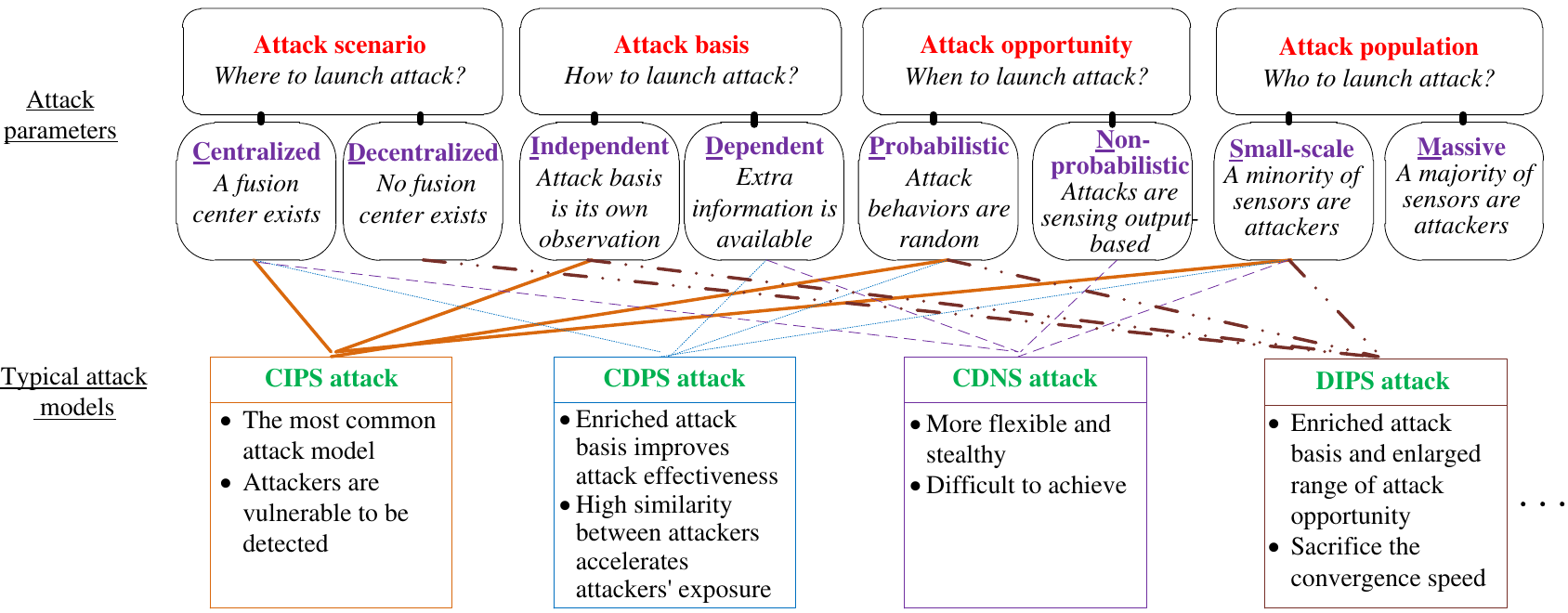}
\caption{Summary of various attack parameters and typical attack models.}
\label{figure-attack-re}
\end{figure}

The Byzantine attack takes advantage of CSS's vulnerability and degrades CSS's performance to achieve its attack objectives through reporting fake data. In the aforementioned, we have introduced four parameters, that is, attack scenario, attack basis, attack opportunity, and attack population, respectively, to illustrate where, how, when, and who to launch the Byzantine attack. In the following, we will briefly summary the main characteristics of every parameter and discuss their roles in typical attack models, just as illustrated in Fig. \ref{figure-attack-re}.

{\emph{1) Attack Scenario: Centralized vs. Decentralized.} Previous studies mostly consider centralized networks in which a reliable fusion center receives and fuses all reports. In centralized networks, global performance and attack performance are easy to be formulated and analyzed. Abundant data resources at the fusion center generally make the attackers vulnerable to be detected. However, the case is quite different in decentralized scenarios, where any user can only communicate with its neighbors and obtain limited information. Moreover, a user's sensing output is always the fusion result of its reports, its neighbors', and (indirectly) its neighbors' neighbors' sensing results via, e.g., an iteration process in the belief propagation algorithm~\cite{Penna_F}. All of these make the situation more complex and create a better shield for attackers in decentralized scenarios.

{\emph{2) Attack Basis: Independent vs. Dependent.} In an independent attack, a malicious user only knows its own observation and has no idea about the real spectrum state and other users' sensing outputs. Nevertheless, the situation becomes better in dependent attacks, where attackers have a better ability of predicting attack results and then achieve better management of attack behaviors. Hence, dependent attacks will be the first choice of a greedy or malicious user, if extra information is available.

{\emph{3) Attack Opportunity: Probabilistic vs. Non-probabilistic.} Attack opportunity is a critical metric influencing malicious users' stealthiness. Specifically, in probabilistic attacks, malicious users decide whether to launch attack or not in each sensing process randomly or with a certain probability. Considering existence of smart attackers aiming to conceal their attack behaviors at the same time of guaranteeing attack performances, non-probabilistic attacks with more flexible setting of attack opportunities have also been investigated, where malicious users decide whether to launch attack or not based on the specified sensing output.

{\emph{4) Attack Population: Small-scale vs. Massive.} In most related work, small-scale attacks are considered, where malicious users are assumed to be in a minority and have limited effects on final decisions. On the other side, the massive attack, where malicious users are assumed to be in a majority, has received quite few attention till now. Nevertheless, as more and more personal devices incorporate into secondary networks, in a small area, the massive attack may occur \cite{Fatemieh_O} and the corresponding attack performance could be much more powerful. Therefore, more research efforts on the massive attack are needed.

{\emph{5) The Roles of Attack Parameters in Typical Attack Models.} Attack parameters determines attack behaviors and the specification and combination of these parameters define various attack models. Provided that each attack parameter has two types as discussed above, it appears that there should be $2^4=16$ attack models by enumerating all the possible combinations the four attack parameters. However, to our best knowledge, so far majority of the potential attack models are rarely investigated in the literature. The commonly studied attack models is summarized as follows:
\begin{itemize}
  \item {The most common attack model is the Centralized Independent Probabilistic Small-Scale (CIPS) attack, which has been widely adopted in the literature. In centralized cooperative sensing scenarios, the number of malicious users is in a minority. Each malicious user independently launches spectrum attack based only on its own observations in the current slot, since it knows nothing about real channel state, other users¡¯ sensing outputs and final fusion results in previous slots. In this case, a common method of choosing an attack opportunity is probabilistic, i.e., to falsify reports with a certain probability at each slot. Generally, there is a tradeoff for each attacker on choosing a proper attack probability to make a balance between destructive and stealthiness. Hence, this kind of attack models is simple and serves as the cornerstone of other attack models.}
  \item Differently, the Centralized Dependent Probabilistic Small-Scale (CDPS) attack considers extra information beyond each attacker¡¯s own observations, such as other users¡¯ sensing results, fusion rules, or even defense strategies, i.e., dependent attack basis. The enriched attack basis enhances attack behaviors. In particular, malicious users may coordinate with each other to falsify data consistently, which eliminates collisions between attackers and poses serious damage on global sensing performances. Nevertheless, in the CDPS attack, the attack opportunity is still probabilistic, which leads to abnormally high correlation of attackers and may finally accelerate malicious users¡¯ exposure.
  \item In the Centralized Dependent Non-probabilistic Small-Scale (CDNS) attack, malicious users consider non-probabilistic attack opportunity to fully exploit attack basis¡¯s timeliness and achieve a more stealthy attack. Specifically, whether to launch attack is related to attack expectancy, which means that attackers will evaluate attack effectiveness before falsifying data. That is, attackers intelligently select attack opportunities, decrease attack risks, and finally improve attack¡¯s persistency.
  \item Compared to attacks in centralized cooperative sensing, Decentralized Independent Probabilistic Small-Scale (DIPS) attack considers decentralized scenarios, in which both attack basis and attack opportunity are enriched. Instead of only its own observations, an attacker can also use the neighbors¡¯ reports as attack basis, which undoubtedly enables it to have an increasingly clear knowledge of other users¡¯ sensing results and real channel states. Simultaneously, the consensus process of decentralized networks makes the attacker able to inject bad data over and over during the algorithm iteration process.
\end{itemize}}

\section{Byzantine Defense}
{In this section, the Byzantine defense is discussed in detail. Firstly, in Section V-A, a classification of current defense algorithms to typical Byzantine attacks is made. Then, according to this classification, in Section V-B, an in-depth tutorial on the state-of-the-art Byzantine defense schemes is given. Finally, we discuss and compare each kind of schemes in terms of strengths and weaknesses in Section V-C.}

Byzantine defense aims to find out Byzantine attackers and mitigate the negative effects of falsified sensing reports on CSS performance. Generally, the defense scheme needs references representing normal behavior to distinguish malicious sensors from honest users. However, a reliable and general reference does not always exist, since most time we have mixed sensing reports from both malicious and honest sensors and we do not know \emph{a priori} who are honest.
As shown in Fig.~\ref{figure-heterogeneous}, from the perspective of data production, sensors' sensing outputs are closely related to wireless channel characteristics (see Section II and Section III-B for details). If the statistical characteristics of wireless channels are identical for all sensors, then one sensor can be a reference for others. This case is considered in most existing related work, which we call the \emph{homogeneous sensing scenario} here. However, this case is only applicable to a small sensing area. When involved with a large sensing area, especially with complex terrain and buildings, it does not work any longer. We call the latter case as the \emph{heterogeneous sensing scenario}. Based on this classification, in Fig. \ref{fig_defense} we sum up previous studies on defense algorithms. The left part of Fig. \ref{fig_defense} is the classification of defense methods, the right part is four kinds of typical attack models (see Section IV-F for details), and the lines between them represent that there are previous studies on defense algorithms to tackle with the corresponding attack models.

\begin{figure}[!t]
\centering
\includegraphics[width=0.5\linewidth]{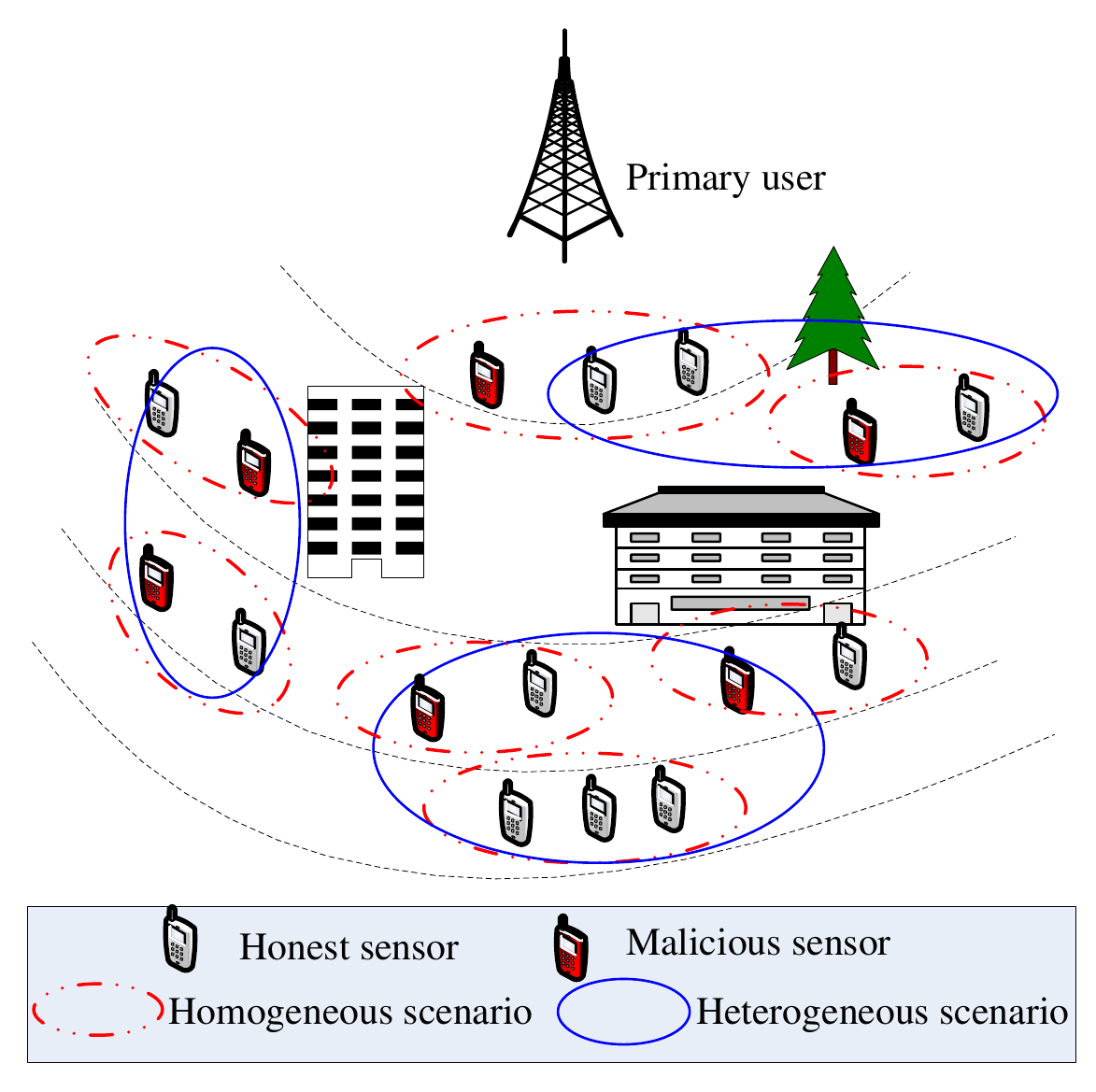}
\caption{Illustration of homogeneous/heterogeneous cooperative spectrum sensing scenarios.}
\label{figure-heterogeneous}
\end{figure}

\begin{figure}[h]
\centering
\includegraphics[width=\linewidth]{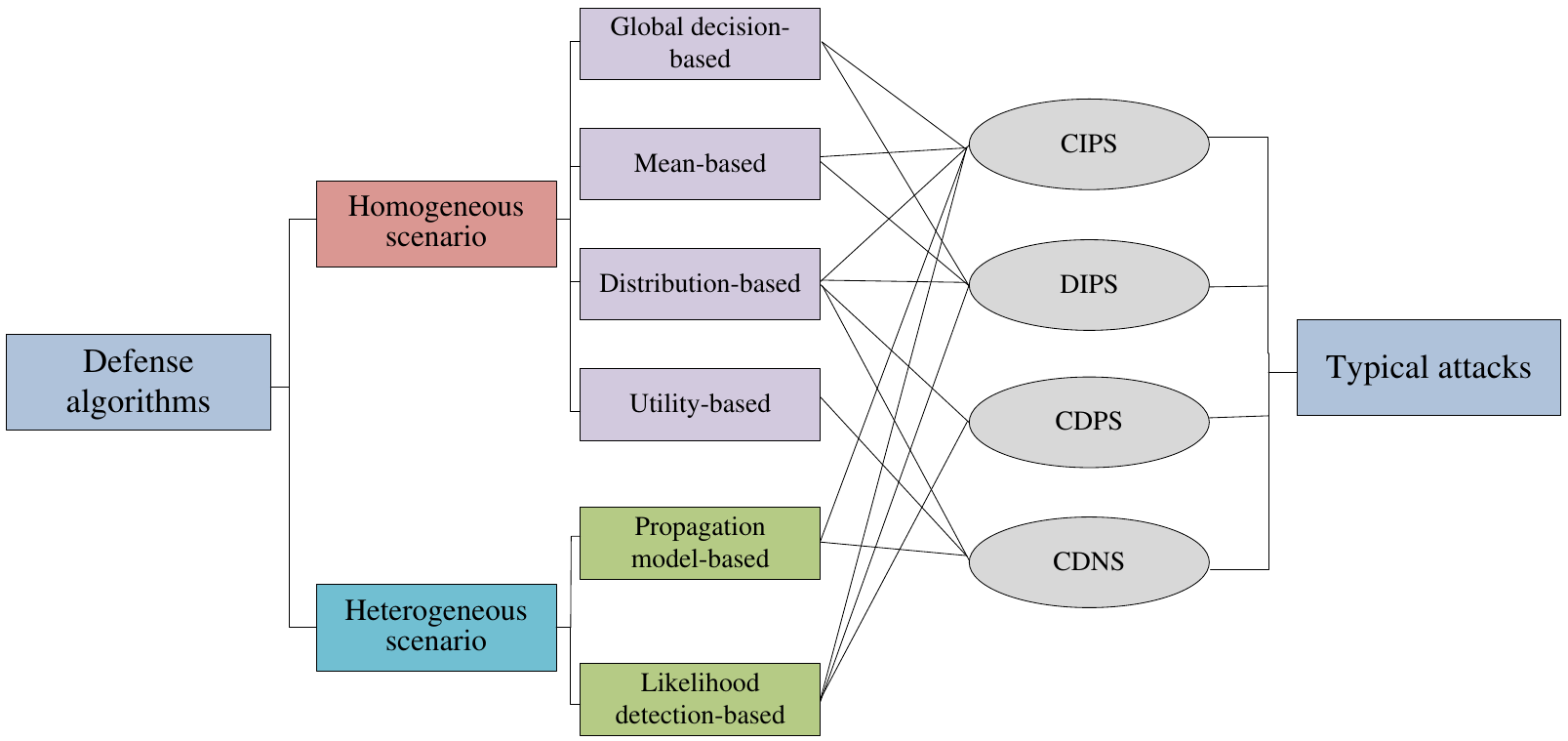}
\caption{Classification of existing defense algorithms against typical Byzantine attacks.}
\label{fig_defense}
\end{figure}

\subsection{Homogeneous Sensing Scenario}
In a homogeneous sensing scenario, all sensors are expected to behave in a similar manner, which makes it appropriate to use a universal standard on behalf of honest ones to evaluate all sensors and find out outliers. The universal references can be concluded as follows:
\begin{itemize}
  \item \emph{Global decision}: A user receives and fuses sensing results from others, compares the fused results with sensors' local decisions, and declares sensors with a low level of consistency as malicious users. In this method, CSS provides the defense scheme with a reference and the defense scheme improves CSS in turn.
  \item \emph{Mean}: The receiver estimates the mean value and variance of reports, and calculates the deviations of a sensor's report and mean value. Although legitimate sensors' observations share some uncertainty, malicious sensors falsify raw observations and report fake results with larger deviations. Here robust estimation of the mean value is critical for the defense performance. Obviously, this method adapts to soft fusion.
  \item \emph{Underlying distribution}: Reports are deployed to analyze reporting behaviors and abnormal sensors are declared as malicious users. Legitimate or honest sensors' reports obey to the identical underlying distribution, while malicious users falsify sensing data and the reports show different distributions. Comparison of reports or metrics extracted from reports can reveal the discrepancy in term of distributions.
  \item \emph{Utility}: Malicious users are assumed as rational individuals in pursuit of maximizing their attack utilities, including incentive and penalty. Malicious users can adapt their behaviors to achieve their objective while the network introduces appropriate penalty and incentives to guide them to stop cheating and report honestly.
\end{itemize}

\subsubsection{Global Decision-Based Defense}
In this scheme, the FC calculates deviations of global decisions ${d^C}(t)$ and local decisions $d(t)$ of SUs to discern Byzantine attackers form honest users. Specifically, the $i$-th SU's deviation at the $t$-th slot can be denoted as
\begin{align}
{\delta _i}(t) = \left\{ \begin{gathered}
  \begin{array}{*{20}{c}}
  {1,}&{{d^C}(t) \ne {d_i}(t),}
\end{array} \hfill \\
  \begin{array}{*{20}{c}}
  {0,}&{{d^C}(t) = {d_i}(t).}
\end{array} \hfill \\
\end{gathered}  \right.
\end{align}

Then, the expected value of deviation ${\delta _i}(t)$, denoted as ${\theta _i}$,  can be used to test whether malicious users exist as follows \cite {Non-parametric,Verikoukis_C}:
\begin{equation}
\left\{ \begin{gathered}
  {A_0}:\;{\theta _1} = {\theta _2} =  \ldots  = {\theta _K}, \hfill \\
  {A_1}:\;{\theta _i} \ne {\theta _j}, ~{\exists}~i \ne j, \hfill \\
\end{gathered}  \right.
\end{equation}
where $A_0$ denotes the null hypothesis that there exist no malicious users, $A_1$ denotes the alternative hypothesis, and $K$ is the number of SUs. In fact, the test above is an ideal case as users' reports are of certain randomness. So the Kruskal-Wallis theory is introduced to perform robust test bearing some minor deviations \cite {Non-parametric}. Reputation is widely introduced into the process of data fusion to mitigate bad effects of malicious falsification on final decisions \cite {enhance_wsprt}\cite {WSPRT}\cite {Nguyen_Thanh}\cite {zhang_x}\cite {kasiri_B}\cite {deleterious}. Essentially, the reputation monotonically increases with the accumulated mean deviations $\frac{1}{T}\sum_{t = 1}^T {{{( - 1)}^{{\delta _i}(t)}}} $. Reputation is deployed in weighted fusion, so SUs with low weights have few effects on final decisions. Hence, CSS provides a defense scheme with a reference, which improves CSS in turn.

Two metrics, the probability of a Byzantine user falsely recognized as an honest one (the probability of false positive), and the probability of an honest user falsely declared as a malicious user (the probability of false negative) are introduce to evaluate defense performances \cite {Rawat_A_S}. Specifically, in \cite {Rawat_A_S}, if one user's reputation metric ${h_i} = \sum_{t = 1}^T {{\delta _i}(t)} $  is greater than a fixed threshold $\eta $, then it is isolated from the data fusion process; otherwise, all users equally participate the fusion. Hence, the two evaluation metrics are denoted as follows \cite {Rawat_A_S}:
\begin{equation}
\begin{gathered}
  P_B^{hon} = P({h_i} > \eta ) = 1 - \sum\limits_{j = \eta  + 1}^T {\left( \begin{gathered}
  T \hfill \\
  j \hfill \\
\end{gathered}  \right)P_B^j{{(1 - {P_B})}^{T - j}}},  \hfill \\
  P_H^{iso} = P({h_i} > \eta ) = \sum\limits_{j = \eta  + 1}^T {\left( \begin{gathered}
  T \hfill \\
  j \hfill \\
\end{gathered}  \right)P_H^j{{(1 - {P_H})}^{T - j}}},  \hfill \\
\end{gathered}
\end{equation}
where ${P_B}$ and ${P_H}$ are the probabilities of inconsistency of a Byzantine or honest SU's reports with the global decision, respectively. Notably, reputation values are closely related to the global decision fusing both honest and malicious users. In order to improve detection reliability, authors in \cite {Przemysaw_Paw} introduce known reliable SUs to make data fusion and determine the credibility of SUs. {In \cite{Safduyu_milcom}, the transmission results on the licensed channel are introduced to enhance detection reliability as a complementarity to global decisions. }

Reliability of global decisions is closely related to fusion rules, the percentage of malicious users, and attack strategies, which determines the method's limited defense performances. Indeed, current defense algorithms based on global decision only consider independent attacks with attack probability as one. Besides, this method mainly applies in hard fusion.

\subsubsection{Mean-Based Defense}
When SUs reports raw observations, legitimate SUs' soft reports are various as the existence of the noise. We need to find a reference on behalf of the legitimate SUs, while global decisions are no longer appropriate. Intuitively, the mean value of reports are selected to find out outliers with large deviations \cite {mobile_ad_hoc}\cite {Bhargava_V_K}\cite {new_technique}\cite {Kaligineedi_P_2010}\cite {dalianli}\cite {ReDiSen}\cite{biologicall_2010}. The involved statistics, in general, are average value $\mu$  and variance value $\sigma$  of reports. The $i$-th reports ${e_i}$' deviation ${\Delta _i}$ from the mean value are exploited as the basis of assigning outlier factors. The basic idea of mean-based outlier detection can be denoted as follows:

\begin{equation}
{\Delta _i} = \frac{{\left| {{e_i} - \mu } \right|}}{\sigma }.
\end{equation}

Notably, the efficiency of this approach is determined by the degree to which the statistics are consistent with the underlying distribution of raw energy values. Unfortunately, it is vulnerable to be degraded by falsified reports from malicious users, which leads to unfavorable skew of the statistic. Hence, robust statistic is the key to efficient and accurate defense. Robust estimation methods range from a simple outlier detection used to identify extreme values \cite {Bhargava_V_K} to the median method \cite {Kaligineedi_P_2010}, to reputation-based method\cite {new_technique}, to bi-weight estimate calculated iteratively \cite {Kaligineedi_P_2010}, to the Orthogonalized Gnanadesikan-Kettenring (OGK) estimate \cite {dalianli}. There are mainly two metrics evaluating robust estimation: breakdown point and computation complexity. The breakdown point is defined as the maximal proportion of outliers that the estimator can tolerate. A fine method ought to have a high breakdown point at the expense of a low computational cost.

After obtaining the two statistic values robustly estimated, two main methods are applied to detect malicious users. One is to assign reputation values to SUs which is closely related to deviations of reports and estimated statistic values \cite {Bhargava_V_K}\cite {dalianli}. The more a SU's reports deviate from the center of observations defined by the two statistics, the lower trust value is assigned to the SU. An alternative approach is to take advantage of fluctuations of robust statistics due to PUs' state handover and the key point can be formulated as \cite {Kaligineedi_P_2010}:
\begin{equation}
\Delta \widehat \mu (k) = \widehat \mu (k) - \widehat \mu (k - 1),
\end{equation}
where $k$ means the sensing slot. As all SUs sense the same licensed channel, an honest SU ought to keep consistent with changes of robust statistics, and if not, it is declared as malicious. This method is similar to change point detection using temporal correlation in \cite {dalianli}.

Mean values mitigate negative effects of uncertainty of observations to some extent and robust estimation sustains part of bad data's influences. But just like decision-based approaches, current related work is limited to be against simple attacks as the CIPS attack or the DIPS attack with the attack probability as 1.

\subsubsection{Underlying Distribution-Based Defense}
In a homogeneous sensing scenario, SUs' observations are expected to obey to one same distribution, i.e., underlying distribution, but due to data falsification, malicious users deviate from the distribution, which can be used to distinguish malicious users from honest ones.
An intuitional method is to compare reports of SUs and evaluate similarity among SUs \cite {catch_me}\cite {Believe_yourself}\cite {Wang_J_2013}\cite {abnormality_icc}\cite {block_outlier}\cite {enhance_DS}. An alternative approach is to extract from reports some metrics representing the distribution \cite {adaptive_byzantine}\cite {HMM_based}\cite {Xiaofan_He}.

The intuitional method takes advantage of different comparison results of SUs' reports to identify malicious ones. In \cite {catch_me}, a robust algorithm called double-sided neighbor distance (DSND) is applied based on the abnormality detection in data mining. It considers two kinds of attack comprehensively: CIPS and CDPS \cite {abnormality_icc}. Distance values of a SU with other SUs are put in ascending order and values of two fixed positions are absorbed and compared with two thresholds, respectively. Considering unreliability of DSDN algorithm in some special cases, a novel method takes advantage of multidimensional correlation to distinguish stealthy malicious users from honest ones \cite {Wang_J_2013}. Besides, a trust node is introduced to enhance the robustness of test \cite {Believe_yourself}.

%Among legitimate SUs, similarity $S_1$ of reports is not weak but not very strong considering limited local detection performance. Among malicious users,
%similarity $S_2$ of reports depends on attack strategies and the same is true for similarity $S_3$ between an honest SU and a malicious user. When attack
%probability is low, $S_2$ and $S_3$ are close to $S_1$, and $S_3$ decrease with attack probability, while independent stealthy attack enhances both $S_2$ and
%$S_3$. Honest users are generally assumed to be in majority.

It is a promising trend to learn malicious SUs' reporting patterns through analyzing reports' distribution. Authors in \cite {HMM_based} consider spectrum state transition probabilities hid in the reports, under the assumption that the true spectrum is Markovian. Due to data falsification, malicious users differ from honest SUs in terms of corresponding Markovian parameters, which is used to separate malicious users from honest users. Instead of directly comparing reports, multiple metrics, indicating the underlying distribution, are considered together to decrease detection time and improve detection accuracy \cite {adaptive_byzantine}\cite {Xiaofan_He}. The method is formulated as \cite {adaptive_byzantine}:
\begin{align}
  \Lambda _i^T = |\widehat \gamma _i^T - {\gamma _H}|
               + \mathop {\min }\limits_{1 \leqslant j \ne i \leqslant N} |{\widehat {{\delta _{ij}}}^T} - {\delta _H}|
                + \mathop {\min }\limits_{1 \leqslant j \ne i \leqslant N} |{\widehat {{\rho _{ij}}}^T} - {\rho _H}|,
\end{align}
with
\begin{align}
&\gamma  = P({u_i} = 1),\nonumber\\
&\delta  = P({u_i} = 1,{u_j} = 1),\nonumber\\
&\rho  = P({u_i} = 0,{u_j} = 0),
\end{align}
where ${\gamma _H},\;{\delta _H},{\rho _H}$ are metrics of honest SUs. Although it relies on a prior knowledge of honest SUs, it cannot only detect malicious users accurately, but also learn behaviors of attackers and assist to make effective fusion.

\subsubsection{Utility-Based Defense}
Intelligent attackers are considered to have the ability of calculating their own utilities and to be in pursuit of maximizing the utilities. Naturally, they are sensitive to penalties and incentives of the system. Hence, the utility-based defense scheme is introduced and the key idea is to guide malicious users to report honestly through adjusting penalty and incentives without identifying attackers directly.

Incentive alleviation is deployed when attackers conduct data falsification in pursuit of access gains denoted as malicious users' utility including no penalties \cite {dos}. The malicious user's utility $\widehat U$ consists of two parts: one is the payoff $U$ accrued by this malicious user being part of CRN, i.e., the utility of a trustful user; while the other is the utility achieved via mis-report. In fact, although falsification brings about the second part of utility, it also decreases the first part. Hence, authors in \cite {dos} intend to design appropriate local threshold and fusion rules to make malicious users realize that falsification does not increase their utility, i.e., $\widehat U < U$, where $U$ denotes its utility when launching no attack. Simulation results show that when the AND fusion rule is applied, the aim can be achieved.

However, the method in \cite {dos} may lead to a large probability of missed detection, but it does not consider negative effects on the PU networks by missed detection. Indeed, harmonious co-existence ought to be established on the basis of valid protection for PUs' spectrum usage and effective exploitation of idle spectrum. Hence, inference incurred by SUs with PUs must be controlled and one feasible approach is collision penalty \cite {coexistence}. That is to say, PU is entitled to charge a collision penalty to all SUs in the network when a collision happens. Authors in \cite {Attack_prevent} consider two kinds of attacks in pursuit of immediate and long-term rewords, respectively.
If the presence of the PU is announced by the FC, legitimate SUs will not transmit data, but Byzantine attackers may consider that their attacks work and the FC is misled, and decide to transmit data. Here is a case in which the licensed channel is occupied by the PU, considering attackers' limited detection performance.
Then the collision happens. Obviously, the FC knows the existence of malicious users. In \cite {Attack_prevent}, the FC punishes the whole SUs through
charging them additional penalty or terminal of cooperative sensing beyond the one charged by the PU.
Through penalties, the FC tries to let attackers realize that the expected utility is always less than that when they behave honestly, considering the gains and penalties together. As a result, a rational attacker, aiming to increase its access gains, will stop falsifying reports. Nevertheless,
an irrational attacker may not care about its own gains, but aims to degrade the whole system's performance \cite {intelligent_attack}. To solve this problem,
spectrum sensing and access mechanisms are jointly considered to prevent the malicious users from falsifying reports \cite {intelligent_attack}.

Utility based defense can protect the networks from suffering from Byzantine attack without detection malicious users. However, the problem is that when malicious users launch attacks, honest users always endure huge and equal penalty but share less gains than malicious users, which is not fair for honest ones. In addition, setting of penalty needs the prior knowledge about attackers which is hard to obtain. This method only adapts to attackers who are intelligence enough to obtain related information, calculate its utility, and adaptively adjust its attack behavior.

\subsection{Heterogeneous Sensing Scenario}
Different from SUs in a homogeneous sensing scenario, SUs, in a heterogeneous scenario, share discrepant detection performances \cite {tan_k}, due to shadowing, path-loss, and multi-path shading, just as illustrated in Fig.3. In addition, SUs may apply different signal detection techniques, which enhance the scenario's heterogeneity as well.
Consequently, evaluation standards in homogeneous scenarios are challenged. In this subsection, defense strategies are introduced to tackle with the Byzantine attack in the heterogeneous scenario, just as conclude below:
\begin{itemize}
  \item \emph{Propagation model based defense}. SUs' observations are closely related to the channel propagation characteristics while data falsification deteriorates the relation. So from the perspective of channel propagation models, the rationality of SUs' reports is evaluated and irrational SUs are identified as malicious users.
  \item \emph{Likelihood detection based defense}. In a heterogeneous adverse scenario, legitimate SUs may explore different SS techniques and suffer from various fading. As there exists great inconsistency among reports, likelihood detection identifies SUs with the largest probability of falsifying data as malicious ones, i.e., ones who most appropriately explants the inconsistency.
\end{itemize}

\subsubsection{Propagation Model-Based Defense}
Legitimate SUs' observations are assumed to be determined by the channel propagation model and PU. Hence, deviations of SUs' sensing results can be weighted by differences of channel characteristics. But falsification of reports deteriorates the relation, which can be taken advantage of to discern malicious users from honest ones. In this subsection, related defense strategies are introduced based on channel propagation models.

Firstly, a simple propagation model is considered where SUs share the same fading factor $\beta$, which is the common ground truth \cite {Mins_a_w}\cite {Zhengrui}, regardless of users' location and its formulation can be simplified as follows:
\begin{equation}
{h_i} = ({d_i}/{d_0}) ^ \beta.
\end{equation}

So, users' sensing performance is determined by the distances $d_i$ between the PU and SUs. Hence, the degree of similarity between SUs is directly reflected on the distance gap $\left| {{d_i} - {d_j}} \right|,\;i \ne j$. Here, the location information of the SUs and PU is the premise of detection; while the location information of SUs can be obtained by GPS and reported to FC and the location information of the PU can be easily obtained via geo-location databases \cite {location} or estimation from reports \cite {mobile_reputation}. One feasible detection approach is to calculate and compare consistency levels between SUs on reports with thresholds defined by corresponding distance gaps \cite {mobile_reputation}\cite {Fastprobe}\cite {distributed_chen}. If a user's consistency degrees with majority of the others don't exceed a threshold, the SU is identified as malicious. An alternative approach is to derive the ground truth from users' reports assisted with location information and identify SUs with abnormal results as malicious ones \cite {Mins_a_w}\cite {Zhengrui}. A framework is proposed, in which the ground truth is estimated based on received reports, and SUs with larger deviations than a threshold are filtered iteratively \cite {Mins_a_w}. However, malicious ones' reports deteriorate the accuracy of estimation, especially when involved with the cooperative attack \cite {Zhengrui}. Hence, in the modified framework \cite {Zhengrui}, a partial decision tree is constructed, improving the robustness of detection mechanisms.

Considering geographical diversity, such as randomly located obstacles, the ground truth does not match with factual situations any more. In other words, the fading factor is not constant any longer and the distance between SUs is not enough to describe the similarity between them any more. A feasible solution is dividing the whole area of interest into multiple small cells \cite {classification}\cite {Fatemieh_O}\cite {Min_a_w}\cite {shadow_fading}\cite {Jana_S}. Every small cell is approximately regarded as a homogeneous sensing scenario \cite {Fatemieh_O} \cite {Jana_S} or a area with the simple propagation model \cite {classification}\cite {Min_a_w}\cite {shadow_fading}. However, there exist relatively large deviations among cells, and appropriate process is needed to obtain fine CSS performance. In \cite {Jana_S}, location reliability is proposed to evaluate path-loss characteristics at different positions, and is considered in the data fusion.

Besides, considering the case that certain cells are dominated by malicious users, cells with abnormal fusion results compared with other cells are lagged as outlier \cite {Fatemieh_O}, or massive malicious users in the cell are filtered out \cite {classification} through reliable training examples. Note that, the scale of the cell deserves carefully selection. On one side, the smaller the cell is, the stronger correlation between reports is. On the other side, estimation randomness has a negative relation with the number of SUs. Hence, here exists a tradeoff in determining the scale of a cell, faced with the Byzantine attack under heterogeneous scenarios.

\subsubsection{Likelihood Detection-Based Defense}
Instead of considering heterogeneity only brought by propagation fading, likelihood detection considers the case that SUs have various spectrum signal detection techniques as well, which increases the uniformity of spectrum sensing. Consequently, spatial deviation cannot tackle with the heterogeneity problem. Nevertheless, from the statistic perspective, SUs are defined with different detection performance denoted by the probabilities of detection and false alarm, which is reflected on reports. Likelihood detection investigates the relation between SUs' detection performances and reports to distinguish the malicious users from the honest SUs. Here a critical factor is whether SUs' probabilities of detection and false alarm are prior known or not. Hence, we firstly introduce defense in the ideal case that the knowledge is prior, and then some primary works are introduced considering the case that the knowledge is not prior.
In the first case, the key idea is to deploy a prior knowledge to calculate the probability of a SU being malicious through history reports named as its suspect level \cite {hit_and_run}\cite {Penna_F}\cite {Sun_Y}\cite {CatchIt}\cite {Pena_F_glo}\cite {cross_layer}\cite{Sun_Y_2010}\cite {online_yao}. Essentially, deviations between SUs' performance known by FC and the real performance are deployed to detect malicious SUs. Specifically, the suspect level of a SU can be obtained by the Bayesian approach, which is calculated as \cite {Sun_Y}:
\begin{equation}
\begin{gathered}
  {\pi _n}(t) \triangleq P({T_n} = M|{\mathcal{F}_t})
   = \frac{{P({\mathcal{F}_t}|{T_n} = M)P({T_n} = M)}}{{\sum\limits_{j = 1}^N {P({\mathcal{F}_t}|{T_j} = M)P({T_j} = M)} }}, \hfill \\
\end{gathered}
\end{equation}
where ${T_n} = M$ denotes the $n$-th SU is malicious, and ${\mathcal{F}_t}$ is all observations of t sensing slots. In order to apply the approach into multiple malicious users' detection, the onion-peeling approach \cite {CatchIt} and belief propagation (BP) on factor graphs \cite {Penna_F} are introduced. ``Onion-peeling" means that the process of identifying malicious users is done iteratively. When the suspicious level of a node is the maximum and beyond a certain threshold, it is identified as a malicious one and excluded from decision-making. Then this process is conducted iteratively until no node has the suspicious level higher than the threshold. The BP algorithm \cite {Penna_F} is used to detect malicious users and estimate their attack probabilities, where it is assumed that the attack model is independent probabilistic attack with the unknown attack probability. Notably, it is a way to learn effectively malicious users' attack behavior and parameters in order to improve defense gains as well as to make reliable and effective data fusion.

The main drawback is that a prior knowledge is hard to obtain and seriously degrades feasibility in turn. Hence, likelihood detection requiring no prior knowledge is done in \cite {Soltanmohammadi}\cite {multinode_sensing}. SUs are assumed to be classified into several classes defined by different operating points, i.e., the detection and false alarm probabilities. The number of classes is assumed to be known, and class parameters are estimated by the iterative expectation maximization (EM) algorithm. Here, the hypothesis and SUs' classification are obtained through taking advantage of their relation with class parameters.

Likelihood detection is powerful to detect the CIPS attack but is vulnerable to the cooperative attack in which malicious users cooperatively falsify reports. Computation complexity requires likelihood detection to detect malicious users in a sequential way, and coordinated reports of malicious users protect them from being found out. Simultaneously, considering the case that intelligent malicious users conduct the specific attacks \cite {catch_me} \cite {abnormality_icc} so that their operating points are same with that of a honest SU, the operating point is not significant to distinguish malicious users from honest ones.

%related works:
%global: \cite {Non-parametric,Verikoukis_C} \cite {Nguyen_Thanh}\cite {enhance_wsprt}\cite {WSPRT}\cite {zhang_x}\cite {kasiri_B}\cite {deleterious} \cite {Rawat_A_S} \cite{selfish_2013} \cite{Safduyu_milcom}
%
%mean: \cite {mobile_ad_hoc}\cite {Bhargava_V_K}\cite {new_technique}\cite {Kaligineedi_P_2010}\cite {dalianli}\cite {ReDiSen}\cite{biologicall_2010}
%
%distribution: \cite {enhance_DS}\cite {catch_me}\cite {Believe_yourself}\cite {Wang_J_2013}\cite {abnormality_icc}\cite {block_outlier} \cite {adaptive_byzantine}\cite {Xiaofan_He}\cite {HMM_based}
%
%utility: \cite {dos} \cite {Attack_prevent}  \cite {intelligent_attack}
%
%prop:\cite {mobile_reputation}\cite {Fastprobe}\cite {distributed_chen}\cite {Mins_a_w}\cite {Zhengrui}
%\cite {classification}\cite {Fatemieh_O}\cite {Min_a_w}\cite {shadow_fading}\cite {Jana_S}
%
%likelihood: \cite {hit_and_run}\cite {Penna_F}\cite {Sun_Y}\cite {CatchIt}\cite {Pena_F_glo}\cite {cross_layer}\cite{Sun_Y_2010}\cite {online_yao}
%\cite {Soltanmohammadi}\cite {multinode_sensing}
\subsection{Further Discussions}

\begin{table}[h]
\begin{center}
\textcolor[rgb]{0.00,0.07,1.00}{\caption{Current study on defense in homogeneous scenarios.}}
\begin{tabular}{|p{0.09\textwidth}|p{0.1\textwidth}|p{0.10\textwidth}|p{0.31\textwidth}|p{0.27\textwidth}|}
\hline
\multirow{2}[6]{*}{\textbf{Classification}} & \multicolumn{2}{|c|}{\textbf{References}} & \multirow{2}[6]{*}{\textbf{Contributions}} & \multirow{2}[6]{*}{\textbf{Strengths and Weaknesses}}\\ \cline{2-3}
& {\textbf{All}} & {\textbf{Representative}}& & \\ \hline
\multirow{2}[6]{*}{\makecell[c]{Global\\ decision-\\based \\method}} &\multirow{2}[6]{*}{\makecell[l]{\cite {Rawat_A_S}\cite {enhance_wsprt} \cite {WSPRT}\\\cite {Nguyen_Thanh}\cite {zhang_x}\cite {Non-parametric}\\ \cite {kasiri_B} \cite{Verikoukis_C} \cite{Safduyu_milcom}\\  \cite{selfish_2013} \cite {deleterious}}}
& 2008: \cite {WSPRT} & \makecell[l]{introduced a reputation-based mechanism\\ to the Sequential Probability Ratio Test\\ (SPRT), which was called WSPRT.}
&\multirow{2}[6]{*}{\makecell[l]{$\bullet$ Simple to conduct and analyze; \\$\bullet$ Low robustness and ineffectiveness \\for dependent attack.}} \\ \cline{3-4}
& & 2011: \cite {Rawat_A_S}& \makecell[l]{analyzed effects of attack population on\\ global performance and proposed a simple\\ algorithm to remove malicious users.}&  \\ \hline
\multirow{2}[6]{*}{\makecell[c]{Mean-\\based \\method}} &\multirow{2}[6]{*}{\makecell[l]{\cite {mobile_ad_hoc}\cite {Bhargava_V_K}\cite {new_technique}\\\cite {Kaligineedi_P_2010}\cite {dalianli}\cite {ReDiSen}\\\cite{biologicall_2010}}}
& 2010: \cite {Kaligineedi_P_2010} & \makecell[l]{proposed robust detection techniques in such\\ cases as constrains of information, and partial\\ known information of PUs or SUs.}  &\multirow{2}[6]{*}{\makecell[l]{$\bullet$ Simple, suitable for soft reports, and \\of  high robustness;\\$\bullet$  Ineffectiveness for dependent attack}} \\ \cline{3-4}
& & 2013: \cite {ReDiSen}& \makecell[l]{proposed a reputation-based distributed\\ sensing scheme to restrict negative effects\\ of attack behaviors.} &  \\\hline
\multirow{2}[6]{*}{\makecell[c]{Distribution-\\based \\method}} &\multirow{2}[6]{*}{\makecell[l]{\cite {catch_me}\cite {adaptive_byzantine}\cite {Believe_yourself}\\\cite {HMM_based}\cite {Wang_J_2013}\cite {abnormality_icc}\\\cite {block_outlier}\cite {Xiaofan_He} \cite {enhance_DS}}}
& 2010: \cite {catch_me} & \makecell[l]{an method was proposed based on abnor-\\mality detection in data mining and detect-\\ability under dependent and independent\\ attack was analyzed.}
&\multirow{2}[6]{*}{\makecell[l]{$\bullet$ Suitable to tackle with dependent \\attackers; \\$\bullet$ High computation complexity and \\long detection delay.}} \\ \cline{3-4}
& & 2013: \cite {HMM_based}& \makecell[l]{proposed to identify malicious users via\\ detecting the difference in two hidden\\ Markov model parameters and investigate\\ soft data fusion to explore falsified data.}&  \\\hline
\multirow{2}[6]{*}{\makecell[c]{Utility-\\based \\method}} &\multirow{2}[6]{*}{\makecell[l]{\cite {Attack_prevent} \cite {dos}  \\ \cite {intelligent_attack}}}
& 2009: \cite {dos} & \makecell[l]{proposed to alleviate the incentive of mis-\\reporting through minimizing the utility\\ difference of truthful and cheating SUs.}  &\multirow{2}[6]{*}{\makecell[l]{$\bullet$ Flexible incentive mechanism and\\fine expectation of the final global\\ performance;\\$\bullet$  Ideal assumption about rational nodes.}} \\ \cline{3-4}
& & 2014: \cite {intelligent_attack}& \makecell[l]{proposed a joint spectrum sensing and access\\ mechanism to thwart the malicious behaviors\\ of rational and irrational intelligent malicious\\ users through incentive compatibility.}&  \\
\hline
\end{tabular}
\end{center}
\label{tab:homo}
\end{table}

\begin{table}[h]
\begin{center}
\textcolor[rgb]{0.00,0.07,1.00}{\caption{Current study on defense in heterogeneous scenarios.}}
\begin{tabular}{|p{0.09\textwidth}|p{0.1\textwidth}|p{0.10\textwidth}|p{0.30\textwidth}|p{0.28\textwidth}|}
\hline
\multirow{2}[6]{*}{\textbf{Classification}} & \multicolumn{2}{|c|}{\textbf{References}} & \multirow{2}[6]{*}{\textbf{Contributions}} & \multirow{2}[6]{*}{\textbf{Strengths and Weaknesses}}\\ \cline{2-3}
& {\textbf{All}} & {\textbf{Representative}}& & \\ \hline
\multirow{3}[6]{*}{\makecell[c]{Propagation\\ model-based\\ method}} &\multirow{3}[6]{*}{\makecell[l]{\cite {classification}\cite {Fatemieh_O}\cite {Min_a_w}\\\cite {shadow_fading}\cite {Mins_a_w}\cite {mobile_reputation}\\\cite {Jana_S}\cite {Fastprobe}\cite {distributed_chen}\\\cite {Zhengrui}}}
& 2011: \cite {shadow_fading} & \makecell[l]{grouped sensors in close proximity as a \\cluster and designed a novel filter based on\\ shadow-fading correlation to identify and\\ penalize abnormal sensing reports.}&\multirow{3}[6]{*}{\makecell[l]{$\bullet$ Factual application environment is \\considered and channel propagation\\ characteristics are fully exploited; \\$\bullet$ Location privacy of SUs is endangered. }} \\ \cline{3-4}
& & 2013: \cite {Jana_S}& \makecell[l]{considered mobile CRNs under attack and\\ proposed location reliability and malicious\\ intention to differentiate poor detection\\ performance from data falsification.}&  \\ \cline{3-4}
& & 2013: \cite {Zhengrui}& \makecell[l]{considered dependent attack, verified the\\ detection uncertainty under the attack, and\\ proposed a modified COI (combinatorial\\ optimization identification) algorithm.}&  \\ \hline
\multirow{2}[6]{*}{\makecell[c]{Likelihood \\detection\\ based \\method}} &\multirow{2}[6]{*}{\makecell[l]{\cite {Soltanmohammadi}\cite {hit_and_run}\cite {Penna_F}\\\cite {Sun_Y}\cite {CatchIt}\cite {Pena_F_glo}\\ \cite {cross_layer}\cite {multinode_sensing}\cite{Sun_Y_2010}\\\cite {online_yao}}}
& 2009: \cite {CatchIt} & \makecell[l]{proposed an onion-peeling approach to\\ remove users with high suspicious levels.}
&\multirow{2}[6]{*}{\makecell[l]{$\bullet$ Ability of tackling with more wrong\\cases; \\$\bullet$ High computation complexity and \\limited to static scenarios.}} \\ \cline{3-4}
& & 2014: \cite {Soltanmohammadi}& \makecell[l]{proposed a novel approach with fast\\ convergence to classify users, estimate their\\ real performance, and detect PUs.}&   \\
\hline
\end{tabular}
\end{center}
\label{tab:Heter}
\end{table}

\subsubsection{Overall Comparison}
{Tables I and II present a summary of various Byzantine defense schemes for homogeneous scenarios and heterogeneous scenarios, respectively. According to the classification in previous two subsections, for each kind of defense strategy, the strengths and weakness of related references are summarized, and the contributions of two or three representative papers are also given in detail. In the following we will provide an overall comparison of existing defense methods.}

In a homogeneous scenario, all users share identical sensing capacities, and if a detection standard is appropriate for a user, it is expected to be appropriate for others. That is, universal standards exist to evaluate all users. Specifically,
\begin{itemize}
  \item The \emph{global decision-based method} takes advantage of similarity between users' binary reports and global decisions to identify malicious users. It is easy to conduct and has low computation complexity. However, the defense performance seriously depends on global sensing which is always deteriorated and even controlled by malicious users. This point basically decides its sensitivity to attack population and probability, i.e., low robustness.
  \item The \emph{mean-based approach} tackles with soft reports, and explores statistics of the reports to detect abnormal data. Different from the global decision-based method, it can obtain higher robustness to attack behaviors \cite{Kaligineedi_P_2010}, which increases complexity absolutely. Nevertheless, correlations between users are always ignored and this kind of approach is generally ineffective for dependent attacks.
  \item The underlying \emph{distribution-based approach} explores similarity levels between sensors and finds out the discrepancy of sensors' reporting patterns. Similarity levels between sensors fine reflect attack models, which can be applied to distinguish stealthy and cooperative attackers. But this method has relatively high computation complexity. In addition, as it wastes a long time to learn users' behaviors, a long detection delay is hard to avoid.
  \item The \emph{utility-based approach} proactively conducts defense from a novel perspective. As it aims to prevent malicious users from falsifying data through controlling users' illegal utilities, it is expected to finally achieve a fine global sensing performance due to all users' honest cooperation. However, lack of detection schemes for malicious users makes legitimate sensors have to sustain considerable loss. Besides, this method assumes that attackers are rational and can adjust their attack behaviors based on available utilities.
\end{itemize}

{Quite differently, there is rarely such a universal standard in a heterogeneous scenario, which determines the discrepancy between defense strategies in the two scenarios. Indeed, defense approaches in homogeneous scenarios cannot be applied in heterogeneous ones unless one critical problem is resolved. In \cite{decoupling}, authors propose to decouple the trustiness and local sensing capability, which is just the core of the problem. Specifically, the paper assumes that SUs report local SNR to FC, and when tackling with low similarity of users' reports to global decisions, users with malicious intention are differentiated from ones with poor local performance. If the two cases are not distinguished, honest users with poor sensing performance will be faced with severe punishment. In fact, cognitive radio is likely to apply in densely populated regions which always have complicated terrains. Hence, the obstacle to defense referred to above, measurements' uncertainty, becomes severe. On one hand, attackers with a cognitive nature may take advantage of it to hide attack behaviors. On the other hand, defense algorithms need tackle with the tradeoff between sensing security and deviation tolerance \cite {deviation_tolerant} in order to guarantee the performance of cooperative spectrum sensing. Clear, defense in heterogeneous scenarios is needed. Specifically,
\begin{itemize}
  \item The \emph{propagation model-based approach} considers the heterogeneous characteristic defined by the channel propagation loss, which is obvious especially in urban environments. It exploits the propagation model to establish the relation between sensors and find out outliers. In consequence, spatial information of users is always required, which may endanger location privacy of users.
  \item The \emph{likelihood detection-based approach} extents application scenarios which are not limited to suffer from propagation effects. This approach identifies sensors with the largest probabilities of falsifying data as malicious ones. However, it is generally limited to static scenarios and not feasible to work in dynamic or mobile CRNs. In addition, high computation complexity of likelihood detection-based approach also limits its application.
\end{itemize}}

\subsubsection{Prevent or Detect}%: Active or Reactive / Active or ReactiveIs the Defense based on
%utility-based schemes belong not only to prevention, but also to active. Prevention based defense happens before attack, and detection based defense exists after attack.
%For a certain defender, its own advantages and disadvantages ought to be firstly analyzed, based on which appropriate defense strategies are chosen. To choose the best appropriate method, except for sensing scenario, it has to decide whether prevention or detection based schemes to select.
{Detection based defense focuses on identifying malicious users and tackling with bad data, including most defense methods above. Differently, prevention based defense has such two meanings: preventive defense which is done before attackers' intrusion, for example, the placement of nodes with high security levels; defense algorithms preventing attackers from launching attack, for example, utility-based defense. Specially, when geographic information is available, some users are located near PUs, under no serious shadowing, and play more important roles in the precess of data fusion. So these users deserve more protection costs. On the other side, effective strategies of adjusting users' utility can decrease attack gains, weaken the motivation to falsify data, and force malicious users to give up falsifying data. }

{In conclusion, the prevention based defense is designed from the view of studying attack behaviors, boosting attack difficulties, and cutting down attack gains. Hence, the direct gain is the decreased probability of attacks existing, which effectively improves the security level. However, the conduction of the two kinds of prevention based defense always need correspondingly one of such two conditions below:
\begin{itemize}
  \item Solid essential foundation, such as assigning expensive sensing equipments more robust to attack. It makes attack harder but also boosts the costs. This method is still nearly empty in defending against SSDF attack, though similar works have been done in secure smart grid \cite{prevention}.
  \item Perfect assumption about attack behaviors, such as selfish and intelligent attackers. It works through decreasing attack gains and compelling attackers to give up to report real data.
\end{itemize}
%That is to say, if one or two conditions are satisfied, prevention based defense is conducted and if no, detection based defense.
Indeed, prevention based defense is, to some extent, an effective complement to detection based defense, which is partly because that the former is not studied as fully as the latter, but also is partly because that the conditions of the former conducting are harder to satisfy.}

\subsubsection{Centralized Networks or Decentralized Networks}%Complement to Decentralized NetworksHow to Apply to Centralized Networks or Decentralized Networks /
{There exist many similarities between centralized networks and decentralized networks, and it is true when defending against Byzantine attack. Indeed, though there exist no FC in decentralized networks, every SU in the networks can be regarded to some extent as a FC with a small population of SUs. In a decentralized network, the secure mechanism is designed for every SU who has the opportunity to possess other sensing results. In the discovery stage, every SU ought to trust itself, evaluate other SUs, and express its recommend \cite {reiforcement_learning}. Hence, in adversarial decentralized networks, each SU doesn't act only as a sensor for spectrum sensing just like in centralized networks, but also acts as a defender against malicious SUs.
Consequently, the detection schemes for identifying malicious users are similar to ones in centralized networks, such as the decision-based defense \cite {zhang_x} \cite{selfish_2013}, the mean-based defense \cite {mobile_ad_hoc} \cite{biologicall_2010}, the underlying distribution based defense \cite {Believe_yourself}\cite {ReDiSen}, and the propagation model based method \cite {distributed_chen}.
However, it is still relatively easy to apply in centralized networks due to availability of integrate data and globally controlling. and some extra problems need to be solved in decentralized networks.}

%in a decentralized network, a SU exchanges sensing results with its neighbors, and short geographical distances between them enable the SU of approximately considering the scenario as a homogeneous one

Firstly, in a decentralized network, a consensus result is reached through an iteration process and during the process, malicious users may inject bad data continuously \cite {Yan_q}\cite {deviation_tolerant} to enhance the attack performance and they can employ known defense strategies to conduct stealthy attack. However, study the iteration process and its convergence characteristic provides a solution to effective defense. As the iteration process precedes, the deviations between the maximum state value and minimum value are monotonically decreasing, until diminishing to zero. Hence, negative effects of continuous malicious injection can be mitigated by adaptively turning down the tolerance for deviations between reports from the neighbors and local updated state value, which can be formulated simply as \cite {Yan_q}: $\lambda (k + 1) = \frac{{Es(k + 1)}}{{Es(k)}}\lambda (k)$, where $Es$ is robust statistic estimation for deviation and $\lambda$ is an adaptive local threshold for detecting malicious behaviors. A similar approach is proposed in \cite {distributed_chen} where a local detection threshold varies with deviations among updated state values in current iteration steps.

Besides, in a centralized network, there exists FC to directly identify and punish a malicious user, while the honest users in decentralized networks have to do it collaboratively. When a SU identifies a neighbor as a malicious one, it needs to broadcast this massage to its neighbors and the receivers determine whether the target user is malicious or not, based on massages from different SUs, which is like a voting process \cite {Yan_q}\cite {distributed_chen} .

\subsubsection{Filter or Keep}%:How to Tackle with Fake Data /
The essential objective of Byzantine defense is to obtain more accurate spectrum sensing results, which is identical to that of CSS and ought to be taken in prior consideration when designing defense algorithms. One problem arises when malicious behaviors are exposed, and it is how to tackle with reports of identified malicious users.

An aggressive approach is to filter or discard all reports from malicious users once they are found out. Absolutely, this approach is conducted simply and can thoroughly eliminate negative effects of identified malicious users. However, when defense costs and risks are involved, the problem becomes complex. On one side, before reliability of defense strategies is verified, the risk of mistakenly regarding honest SUs as malicious ones exists, and discarding them will increase defense costs and worsen spectrum sensing performances. On the other side, even reports from malicious users play a positive role in the process of CSS, considering the case that reports are lightly falsified, for example, malicious users launch attacks with low attack probabilities \cite {Penna_F}.

An alternative method is the reputation-based scheme in which weight values are assigned to users based on suspect levels so that the system can differentiate SUs' influence on final decisions \cite {WSPRT}. This method avoids potential defense costs in a certain degree and keeps part of application value of users whose behaviors are not too bad. However, the assignment of reputations is heuristic and lack of solid theoretical proof, which means that bad data is not dig out really. Simultaneously, the networks may endure sustaining damages on detection performance bad data produces. Instead, information theory tells that only if reports from malicious users are thoroughly random, the reports convey no information. Taking advantage of Byzantine attackers' reports is studied in \cite {adaptive_byzantine,HMM_based} and the key idea is to obtain real sensing results from falsified reports based on attack parameters. But effective and reliable exploitation must be established on the premise of a prior or learned knowledge of accurate attack strategies which is not easy to obtain. In particular, the authors in \cite{IEEE-TCOM-Ding} take advantage of the under-utilization of licensed spectrum bands and the sparsity of nonzero abnormal data to robustly cleanse out the nonzero abnormal data component. Due to full exploitation of sensing data, the final performance is favorable under no prior knowledge.

In conclusion, the selection between filtering and keeping can be regarded as a tradeoff between defense risks and spectrum sensing performance gains. The optimal strategies are ones achieving the best detection performance at the cost of low defense risks.

\section{Future Research Directions}
Although a number of studies have been done on the Byzantine attack and defense for CSS in CRNs, this research topic has a great potential for future investigation. Actually, there are still a number of challenges unsolved and open issues waiting for solutions. Based on the comprehensive survey and tutorial on the previous work above, in this section we first present a tree illustration of potential future research directions in Fig. \ref{Fig-future-work}, {including one tree trunk, three tree limbs, and several tree branches. The tree trunk represent the interactive game between attack and defense. Three tree limbs respectively represents model, method and optimization of attack and defense, while each tree branch denotes one specific research direction under the corresponding tree limb. In the following, we provide the detailed explanations on each research direction, sequentially.}

\begin{figure}[h]
\centering
\includegraphics[width=0.6\linewidth]{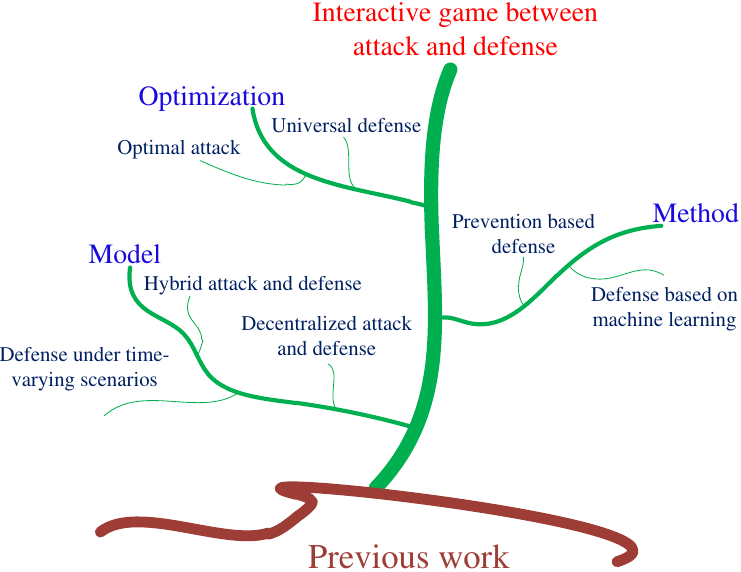}
\caption{A tree illustration of future research directions.}
\label{Fig-future-work}
\end{figure}

\subsection{Interactive Game between Attack and Defense}

\begin{figure*}[h]
\centering
\includegraphics[width=0.9\linewidth]{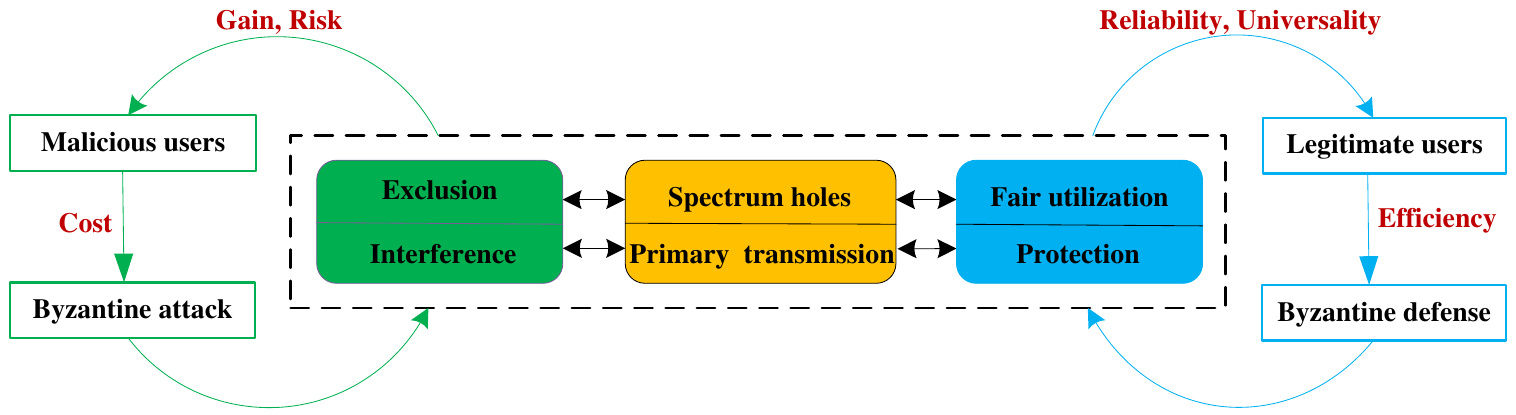}
\caption{Interactive game between attack and defense.}
\label{Fig-architeture}
\end{figure*}

As illustrated in Fig. \ref{Fig-architeture}, spectrum holes and primary transmission can be regarded as system resources. Byzantine attackers and defenders are in pursuit of favorable assignment of resources to themselves. Previously, in \cite {coalitional_game}, a coalitional game is formulated for attackers to maximize the number of invaded channels in a distributed manner. In~\cite {no_regret}, from the perspective of defenders, the nature is assumed to be a player to develop a game-theoretical point of view. However, the research on the interactive game between the Byzantine attack and defense is still in infancy. Specifically, from the perspective of Byzantine attackers, when launching the attack, Byzantine users have to take into account of the factors of attack gain, attack cost and attack risk, together. On the other hand, from the perspective of Byzantine defenders, there are also three aspects deserving consideration: defense reliability, defense efficiency and defense universality. The Byzantine attack and defense are mutually coupled from each other. Attackers need to adjust their strategies to keep their negative effects on final decisions and avoid defenders' detection, while defenders have to learn and analyze attack behaviors and designs effective defense rules. Indeed, attack and defense ought to be considered together.

\subsection{Model Assumption}
\subsubsection{Decentralized Attack and Defense}
It is still quite a challenging direction to research the Byzantine attack and defense in decentralized networks. Decentralized networks, such as mobile ad hoc networks, have found increasingly wide applications in mobile social networking, wireless internet of things, and machine-to-machine communications, etc. Traditionally, comparing with centralized networks, decentralized networks are well recognized as of low communication overhead among only local nodes, and robustness to single node or link failure. Nevertheless, compared to attackers in centralized networks, attackers in decentralized network take more advantages such as a prior knowledge of the fusion rule, defense algorithms, and neighbors' state information. These advantages enable attackers to flexibly adjust attack behaviors. Besides, continuous injection of vicious data during the iteration process of distributed algorithms (e.g., consensus, diffusion, and belief propagation) can change the consensus results incrementally and stealthily. Consequently, the convergence time will increase and the convergence result will deviate from the ground-truth as continuous injections during the iteration process.
\subsubsection{Defense under Time-Varying Scenarios}
It will be a common thing in CRNs that the scenarios are time-varying due to, e.g., the mobility of PUs and/or SUs, the variation of attack behaviors and PU activities, etc. Correspondingly, efficient defense schemes should dynamically adapt to time-varying scenarios. Specifically, in mobile CRNs, every SU's sensing environment is time-varying and when evaluating the confidence level of a SU's sensing report, location diversity must be taken into consideration~\cite{mobile_reputation,Jana_S}. Intelligent attackers will take advantage of the mobility characteristic to launch more intangible and powerful attack, which deserves considerable attention and is a challenging problem for defense. Besides, both PUs and SUs can be mobile and the problem can be more complex. This requires defense schemes to be location-aware.
\subsubsection{Hybrid Attack and Defense}
Various types of hybrid attacks are emerging:
\begin{itemize}
  \item Hybridization of various Byzantine attack strategies. Malfunctioning users and intentional attackers may show different behaviors and attackers belong to different sets, which may conduct dependent or independent attack, respectively.
  \item Hybridization of various attacks in physical layer, e.g., Byzantine attacks can be combined with the primary user emulation (PUE) attack \cite{JSAC-BP,Peng_Q_tvt}.
  \item Hybridization of cross-layer attacks, e.g., the physical layer Byzantine attack can be combined with other attacks in higher layers, such as a small-backoff window attack, and the so-called lion attack \cite{survey-security} \cite {cross_layer}.
\end{itemize}

In the latter two types above, Byzantine attack coordinates with other attacks to achieve powerful performances, but how to cooperate and to the launch Byzantine attack well still has no good solution. On the other side, defense algorithms need to be designed against confusing from other attackers and it is possible to integrate the authentication from higher layers into the MAC-PHY layer \cite{survey-security}.

\subsection{Method}
\subsubsection{Machine Learning based Defense}
One critical part of defense against the Byzantine attack is to analyze data and dig out useful information related with users' behaviors, while machine learning provides powerful data mining tools to achieve this task. In fact, there are already several preliminary studies using various machine learning methods, such as clustering algorithms (see, e.g.,~\cite{kernel_learning,catch_me}) and pattern extraction algorithms (see, e.g., \cite{IEEE-TCOM-Ding,Wang_J_2013}), to defense against the Byzantine attack. However, the theory of machine learning itself has gained increasing research interest recently~(see, e.g.,~\cite{ML-Book-2014,ML-Book2-2014,ML-JMLR}), more efforts are needed to bridge the theory advances in machine learning and the specific defense algorithm designs.
\subsubsection{Prevention based Defense}
Considering the increasingly large and open networks, security problems become more and more serious and have received growing attention. Traditional detection based defense is reactive and may not satisfy various needs of security, while prevention based defense provides the problems with a more flexible and proactive solution. Specifically, prevention based defense, a promising kind of defense schemes, is regarded as one aiming to increase difficulty and risks of attack behaviors and actively develop the defender's advantages, which may contain such two factors:
\begin{itemize}
  \item Prevention before attack. It aims to make a targeted reinforcement considering networks' vulnerability, which will enhance the system's inborn robustness to attack. For example, selectively improve critical nodes' security levels. In consequence, the attack cost is increased while attack effectiveness declines.
  \item Appropriate counterattack. First of all, offending is coordinated with traditional defense schemes including bad data detection. After a certain period, the knowledge of malicious users' behaviors are gradually obtained and the network develops the capacity of launching counterattack. Here, the counterattack means that users identified as malicious ought to be in certain punishment, which will improve users' responsibility widely neglected by users due to the openness of the underlying protocols.
\end{itemize}

\subsection{Optimization}
\subsubsection{Universal Defense}
So far there have been many attack models for the Byzantine attack in CSS and each attack model has several attack parameters, and it is expected that more attack models and parameters are emerging. Generally, majority of existing defense algorithms have been designed for specific attack models and perform well only under specific attack parameters. To our best knowledge, the work in~\cite{IEEE-TCOM-Ding} represents one of the first tries to propose a generalized approach for Byzantine attack modeling, and the data cleansing-based defense scheme developed in~\cite{IEEE-TCOM-Ding} has been shown to perform well under various attack parameters via computer simulations. However, theoretically, there are still no clear answers to the following issues: What characteristics should a universal defense scheme have? How could we design a universal defense scheme? And more practically, how could we employ the existing defense algorithms in the literature to universally defend the already known kinds of attack behaviors? One key challenge is that: due to the spear-and-shield relation between the Byzantine attack and defense, the defense system cannot obtain cooperation from the hostile attackers, and thus do not know the specific attack model and parameters before deploying the proper defense algorithm. Consequently, attack pattern/model recognition and attack parameter estimation are suggested as two key techniques for the design of a practical and universal defense scheme.

\subsubsection{Optimal Attack}
The optimal attack is considered to have the ability to optimize attack performances of Byzantine attacker(s), at the same time, generally serves as the worst case for the design of a robust defense system. For every rational Byzantine attacker, the possible goals to launch an attack mainly include: i) maximization of the attack gain (i.e., the destructive to CSS), ii) minimization of the attack cost (e.g., time and energy consumption), and iii) minimization of the attack risk (e.g., being captured and punished). Generally, there are interesting tradeoffs among these interactive goals.

From the perspective of Byzantine attackers, apart from attack scenarios, the other three parameters can be chosen by Byzantine users. Attack population can be increased by introducing more malicious users, attack basis can be obtained by enhancing the coordination between malicious uses, and attack probability can also be optimized to decrease the CSS performance. However, although Byzantine users benefit from the increased attack population and enriched attack basis, they have to pay for the companying attack cost, such as attack time, attack resources, and communication costs \cite {tree_based}\cite {sybil_attack}\cite {coalitional_game}. Moreover, they have to consider whether or not to attack more frequently \cite {distributed_bayesian}\cite {optimal_distributed}, i.e., to achieve more instant attack gain but endure more risk of being detected \cite  {paper,catch_me,Sun_Y_2010}. At the same time, in the dependent attack, the coordinated attack leads to high correlation between attackers, which may accelerate the exposure of attack behaviors and increase attack risks. All of these deserve more attention when designing the optimal attack strategies.

%\subsection{Performance Metrics Design}

\section{Conclusions and discussions}
This paper has provided a comprehensive overview of the studies on Byzantine attack and defense for CSS in CRNs. To begin with, we analyzed the vulnerability of CSS to the Byzantine attack, the obstacle to defense and the spear-and shield relation between the Byzantine attack and its corresponding defense. Next, we highlighted the diversity of Byzantine attack models from four aspects as where, who, how, and when to launch attacks in details. Then, the state-of-the-art Byzantine defense schemes were elaborated from the perspective of homogeneous and heterogeneous sensing scenarios, respectively. Finally, we depicted the potential future research directions. In a nutshell, the Byzantine attack in CSS is one key adversary to the success of CRNs. More efforts are needed to tackle with the unsolved research challenges on the Byzantine attack and defense in CRNs.

{Furthermore, it is well known that the term ``Byzantine attack'' stems from the Byzantine general problem in
which loyal generals of Byzantine army try to agree upon a common battle plan on in the presence of one or
more traitors\cite{Byzantine_generals}. Therefore, the Byzantine attack generally exists in collaborative scenarios with a group of participators. Actually, besides CSS, the research on Byzantine attack in many other scenarios does exist, such as collaborative localization in wireless sensor networks \cite{localization}, collaborative routing in multi-hop wireless networks \cite{routing}, and collaborative state estimation in smart grid \cite{survey_smgrid,prevention}, etc. We envision that the topic of Byzantine attack and defense is a fruitful research area.
}
%\textcolor[rgb]{0.00,0.00,1.00}{In both scenarios, there exist many data resources, based on which final values are obtained. But what is different is the sensing objectives and data processing measures. Instead of PU's binary state, the state in smart grid is soft. Hence, falsified data, in CRNs, makes final decisions flipped, while bad data, in smart grid, enlarges estimation errors. In smart grid, observations are reflection of several system states through certain network parameter. Instead of obtaining final values through data fusion in CRNs, system states are estimated through coordination of observations with power flow metrics.

%This leads to a critical difference on attack behaviors. In smart grid, the widely acknowledged way of stealthy attack is injection of highly structured bad data that conforms to the network topology and some particular physical laws, which is apparently different from stealthy attack in CRNs referred to above.

%Simultaneously, the premise of final state values' availability in smart grid is that enough observations are reported due to the special way of obtaining final values. It means that in smart grid, observations is not discarded as much as possible. Hence, prevention based defense schemes have received great attention, in which sensors in critical positions are protected from being intruded.}

%\IEEEtriggeratref{5}
\section*{Acknowledgements}
We thank the editor and anonymous reviewers for their precious time and efforts in reviewing this paper and
providing us many helpful comments and constructive suggestions, which have helped us improved the quality of
this paper significantly.

%-----------------------------------------------------------------------------------------------------------------------------------------------------
\begin{IEEEbiography}[{\includegraphics[width=1in,height=1.25in,clip,keepaspectratio]{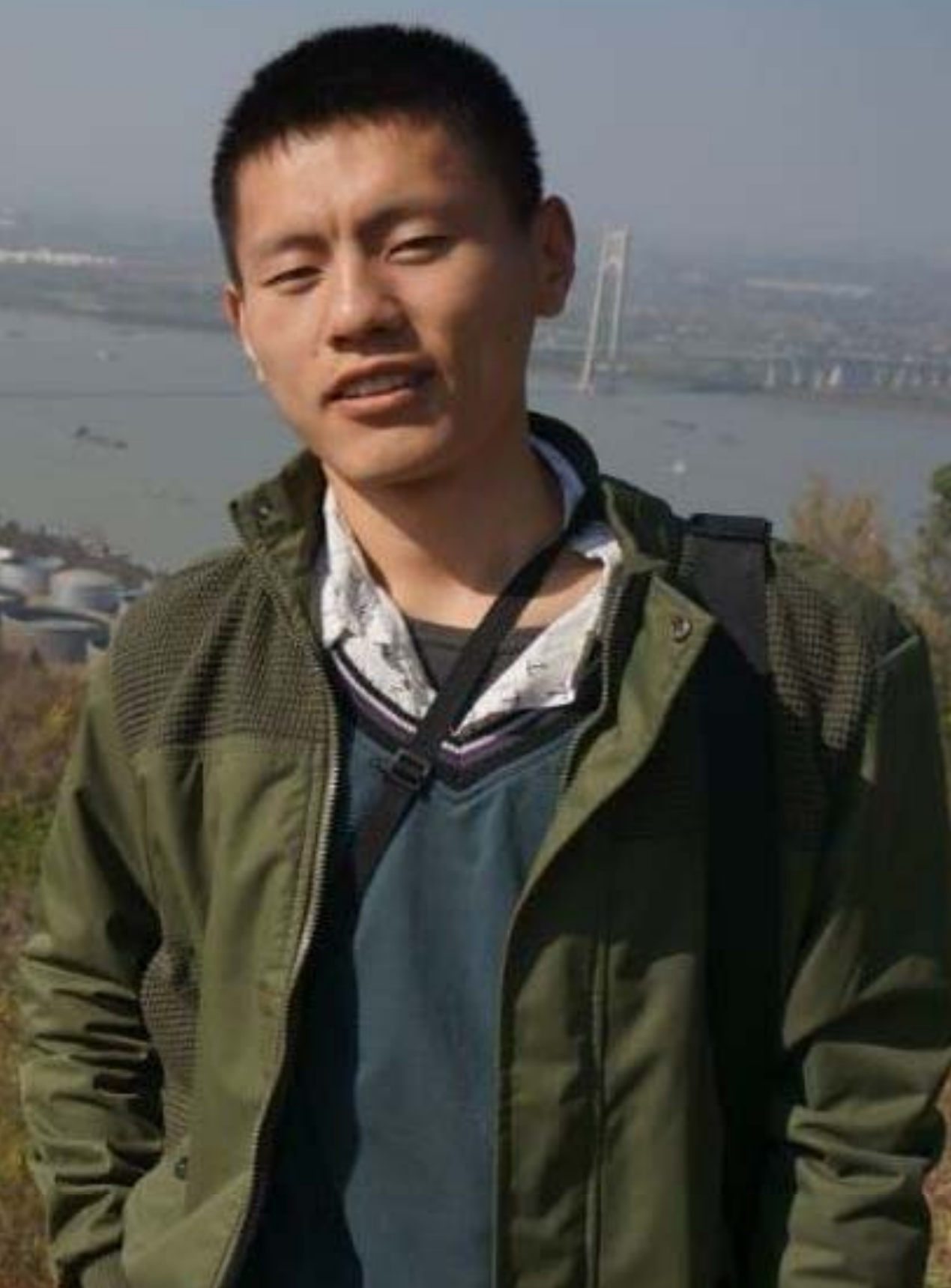}}]{Linyuan Zhang}
received his B.S. degree (with honors) in electronic engineering from Inner Mongolia University, Hohhot, China, in 2012. He is currently pursuing his M.S. degree in communications and information system in College of Communications Engineering, PLA University of Science and Technology. His research interests are wireless communications and cognitive radio networks.
\end{IEEEbiography}
%-----------------------------------------------------------------------------------------------------------------------------------------------------

%-----------------------------------------------------------------------------------------------------------------------------------------------------
\begin{IEEEbiography}[{\includegraphics[width=1in,height=1.25in,clip,keepaspectratio]{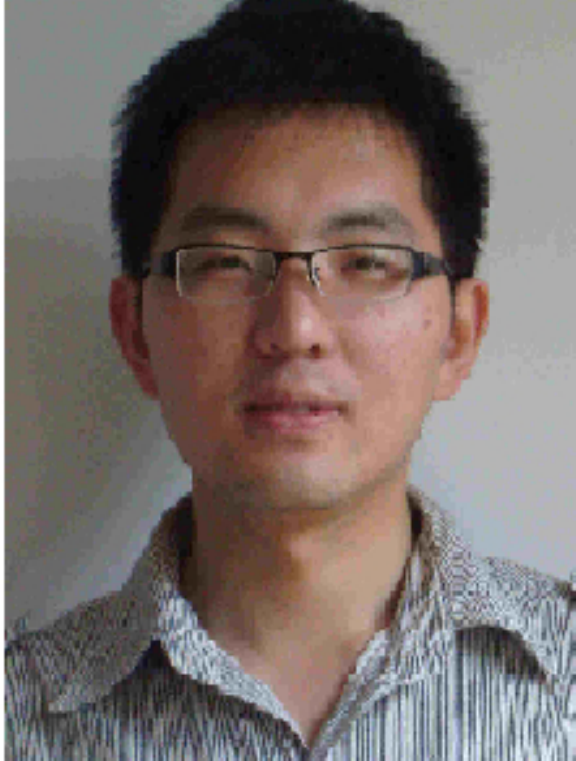}}]{Guoru Ding}
received his B.S. degree in electrical engineering from Xidian University, Xi'an, China, in 2008 and his Ph.D. degree in communications and information systems in College of Communications Engineering, PLA University of Science and Technology, Nanjing, China, in 2014. His research interests include wireless security, cognitive radio networks, machine learning, and big data analytics over wireless networks.

He currently serves as a regular reviewer for 10+ international journals, including IEEE Signal Processing Magazine, IEEE Communications Magazine, IEEE Transactions on Signal Processing, IEEE Transactions on Communications, and IEEE Transactions on Wireless Communications, etc. He is member of the Editorial Boards of KSII Transactions on Internet and Information Systems. He has acted as Technical Program Committees (TPC) members for a number of international conferences, including the IEEE Global Communications Conference (GLOBECOM), IEEE International Conference on Communications (ICC), and IEEE Vehicular Technology Conference (VTC), etc. He was a recipient of the Best Paper Awards from IEEE VTC 2014-Fall and IEEE WCSP 2009. He actively participates in international standardization association IEEE DySPAN Standards Committee and acts as the Secretary of IEEE 1900.6 and one of the voting members both in IEEE 1900.7 and IEEE 1900.6.
\end{IEEEbiography}
%-----------------------------------------------------------------------------------------------------------------------------------------------------

%-----------------------------------------------------------------------------------------------------------------------------------------------------
\begin{IEEEbiography}[{\includegraphics[width=1in,height=1.25in,clip,keepaspectratio]{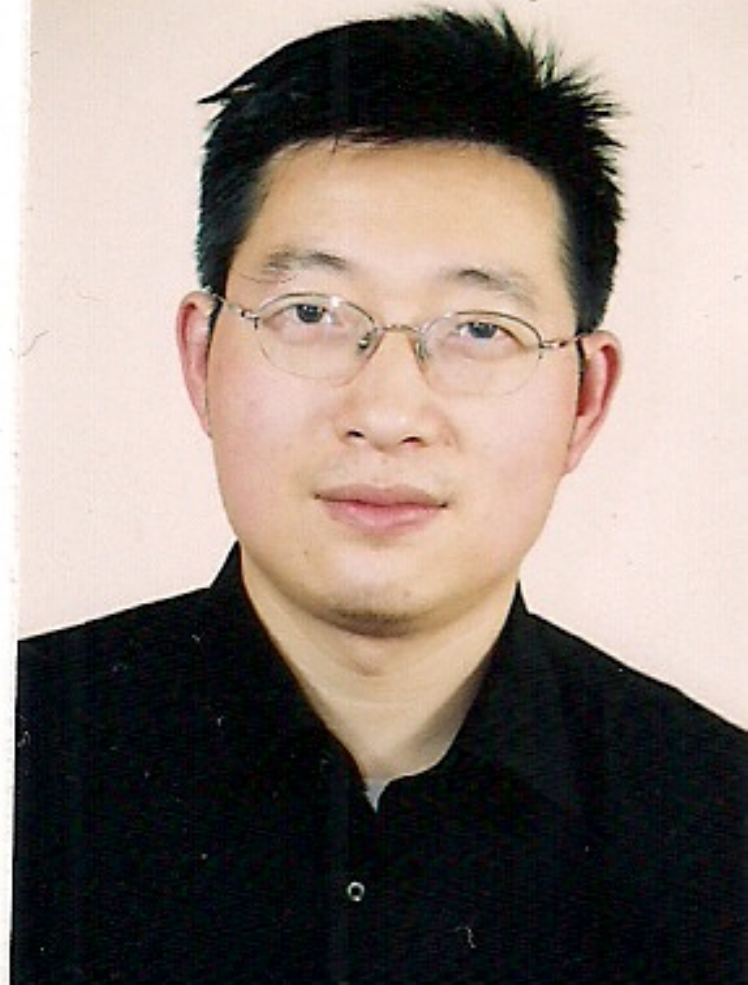}}]{Qihui Wu}
received his B.S. degree in communications engineering, M.S. degree and Ph.D. degree in communications and information systems from Institute of Communications Engineering, Nanjing, China, in 1994, 1997 and 2000, respectively. From 2003 to 2005, he was a Postdoctoral Research Associate at Southeast University, Nanjing, China. From 2005 to 2007, he was an Associate Professor with the Institute of Communications Engineering, PLA University of Science and Technology, Nanjing, China, where he is currently a Full Professor. From March 2011 to September 2011, he was an Advanced Visiting Scholar in Stevens Institute of Technology, Hoboken, USA.

Dr. Wu's current research interests span the areas of wireless communications and statistical signal processing, with emphasis on system design of software defined radio, cognitive radio, and smart radio.
\end{IEEEbiography}
%-----------------------------------------------------------------------------------------------------------------------------------------------------

%-----------------------------------------------------------------------------------------------------------------------------------------------------
\begin{IEEEbiography}[{\includegraphics[width=1in,height=1.25in,clip,keepaspectratio]{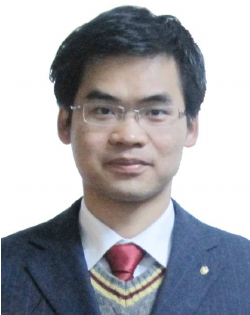}}]{Yulong Zou}
(S'07-M'12-SM'13) is a Full Professor at the Nanjing University of Posts and Telecommunications (NUPT), Nanjing, China. He received the B.Eng. degree in Information Engineering from NUPT, Nanjing, China, in July 2006, the first Ph.D. degree in Electrical Engineering from the Stevens Institute of Technology, New Jersey, United States, in May 2012, and the second Ph.D. degree in Signal and Information Processing from NUPT, Nanjing, China, in July 2012. His research interests span a wide range of topics in wireless communications and signal processing, including the cooperative communications, cognitive radio, wireless security, and energy-efficient communications. Dr. Zou is currently serving as an editor for the IEEE Communications Surveys and Tutorials, IEEE Communications Letters, EURASIP Journal on Advances in Signal Processing, and KSII Transactions on Internet and Information Systems. He has also served as the lead guest editor for a special issue on ``Security Challenges and Issues in Cognitive Radio Networks'' in the EURASIP Journal on Advances in Signal Processing. In addition, he has acted as symposium chairs, session chairs, and TPC members for a number of IEEE sponsored conferences, including the IEEE Wireless Communications and Networking Conference (WCNC), IEEE Global Communications Conference (GLOBECOM), IEEE International Conference on Communications (ICC), IEEE Vehicular Technology Conference (VTC), International Conference on Communications in China (ICCC), and so on.
\end{IEEEbiography}
%-----------------------------------------------------------------------------------------------------------------------------------------------------

%-----------------------------------------------------------------------------------------------------------------------------------------------------
\begin{IEEEbiography}[{\includegraphics[width=1in,height=1.5in,clip,keepaspectratio]{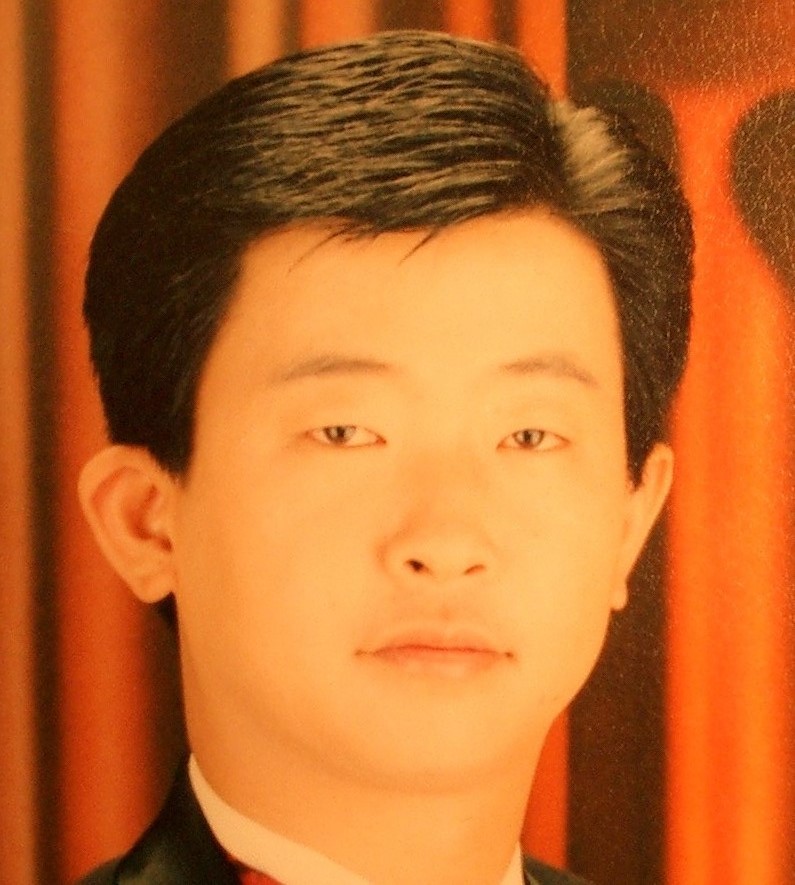}}]{Zhu Han}
(S'01-M'04-SM'09-F'14) received the B.S. degree in electronic engineering from Tsinghua University, in 1997, and the M.S. and Ph.D. degrees in electrical engineering from the University of Maryland, College Park, in 1999 and 2003, respectively.

From 2000 to 2002, he was an R$\&$D Engineer of JDSU, Germantown, Maryland. From 2003 to 2006, he was a Research Associate at the University of Maryland. From 2006 to 2008, he was an assistant professor in Boise State University, Idaho. Currently, he is an Associate Professor in Electrical and Computer Engineering Department at the University of Houston, Texas. His research interests include wireless resource allocation and management, wireless communications and networking, game theory, wireless multimedia, security, and smart grid communication. Dr. Han is an Associate Editor of IEEE Transactions on Wireless Communications since 2010. Dr. Han is the winner of IEEE Fred W. Ellersick Prize 2011. Dr. Han is an NSF CAREER award recipient 2010. Dr. Han is IEEE Distinguished lecturer since 2015.
\end{IEEEbiography}
%-----------------------------------------------------------------------------------------------------------------------------------------------------

%-----------------------------------------------------------------------------------------------------------------------------------------------------
\begin{IEEEbiography}[{\includegraphics[width=1in,height=1.25in,clip,keepaspectratio]{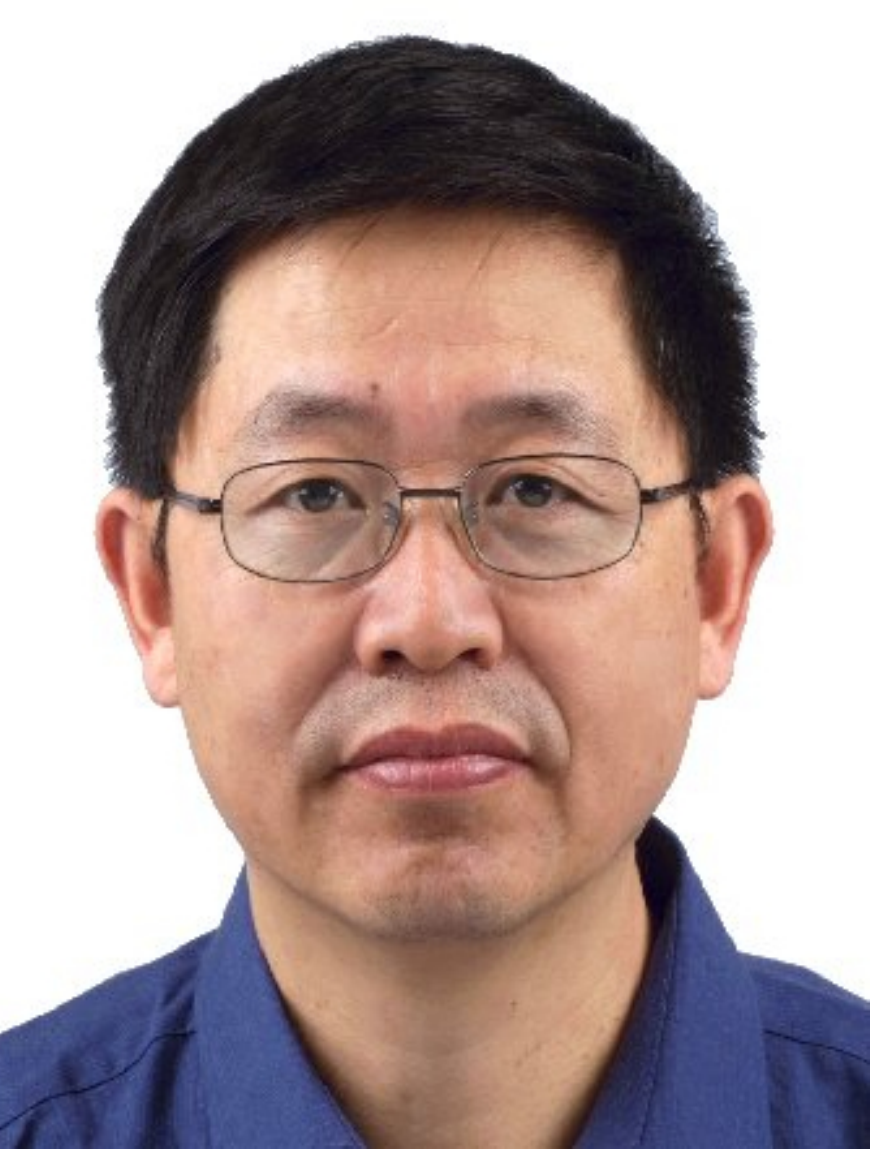}}]{Jinlong Wang}
received his B.S. degree in wireless communications, M.S. degree and Ph.D. degree in communications and electronic systems from Institute of Communications Engineering, Nanjing, China, in 1983, 1986 and 1992, respectively.

He is currently a Chair Professor at PLA University of Science and Technology, Nanjing, China. He is also the co-chair of IEEE Nanjing Section. He has published widely in the areas of signal processing for wireless communications and networking. His current research interests include soft defined radio, cognitive radio, and green wireless communication systems.
\end{IEEEbiography}
%-----------------------------------------------------------------------------------------------------------------------------------------------------

\end{document}